\documentclass[aps,prd,preprintnumbers,superscriptaddress,showpacs,showkeys,floatfix,amsmath,amssymb,amsfonts,twocolumn,10pt]{revtex4-2}

\usepackage{graphicx}
\usepackage{multirow}
\usepackage{makecell}
\usepackage{xparse} 
\usepackage{float}
\usepackage{xcolor}

\NewDocumentCommand\Nf{mgg}{N\textsubscript{f}=#1\IfNoValueTF{#2}{}{+#2}\IfNoValueTF{#3}{}{+#3}}
\NewDocumentCommand\vol{mg}{#1\textsuperscript{3}\IfNoValueTF{#2}{}{×#2}}

\newcommand{\tins}{t_\mathrm{ins}}
\newcommand{\xins}{x_\mathrm{ins}}
\newcommand{\vxins}{\vec{x}_\mathrm{ins}}

\usepackage[normalem]{ulem}

\usepackage{hyperref}
\usepackage{slashed}
\begin{document}

\title{Nucleon strange electromagnetic form factors from \Nf{2}{1}{1} lattice QCD}

\author{Constantia Alexandrou} \affiliation{Department of Physics, University of Cyprus, P.O. Box 20537, 1678 Nicosia, Cyprus}\affiliation{Computation-based Science and Technology Research Center, The Cyprus Institute, 20 Kavafi Str., Nicosia 2121, Cyprus}
\author{Simone Bacchio} \affiliation{Computation-based Science and Technology Research Center, The Cyprus Institute, 20 Kavafi Str., Nicosia 2121, Cyprus}
\author{Mathis Bode} \affiliation{J\"{u}lich Supercomputing Centre, Forschungzentrum Jülich, D-52425 Jülich, Germany}
\author{Jacob Finkenrath} \affiliation{Department of Physics, Bergische Universit\"{a}t Wuppertal,
Gaußstr. 20, 42119 Wuppertal, Germany}
\author{Andreas Herten} \affiliation{J\"{u}lich Supercomputing Centre, Forschungzentrum Jülich, D-52425 Jülich, Germany}
\author{Christos Iona} \affiliation{Department of Physics, University of Cyprus, P.O. Box 20537, 1678 Nicosia, Cyprus}\affiliation{Computation-based Science and Technology Research Center, The Cyprus Institute, 20 Kavafi Str., Nicosia 2121, Cyprus}
\author{Giannis Koutsou} \affiliation{Computation-based Science and Technology Research Center, The Cyprus Institute, 20 Kavafi Str., Nicosia 2121, Cyprus}
\author{Ferenc Pittler} \affiliation{Computation-based Science and Technology Research Center, The Cyprus Institute, 20 Kavafi Str., Nicosia 2121, Cyprus}
\author{Bhavna Prasad} \affiliation{Computation-based Science and Technology Research Center, The Cyprus Institute, 20 Kavafi Str., Nicosia 2121, Cyprus}
\author{Gregoris Spanoudes}\affiliation{Department of Physics, University of Cyprus, P.O. Box 20537, 1678 Nicosia, Cyprus}

\date{\today}

\begin{abstract}
	We present the nucleon strange electromagnetic form factors
	using four lattice QCD ensembles  with \Nf{2}{1}{1} twisted mass
	clover-improved fermions and quark masses tuned to approximately their
	physical values. The four ensembles have similar physical
	volume and lattice spacings of $a=0.080$~fm, $0.068$~fm,
	$0.057$~fm and~$0.049$ fm allowing us to take the continuum
	limit directly at the physical pion mass point. We compute nucleon three-point
	correlation functions with high statistics, where   the
	disconnected fermion loops are evaluated stochastically with
	spin-color dilution and hierarchical probing. We find non-zero values for both electric and magnetic form factors. We extract the
	strange electric and magnetic radii, as well as the strange
	magnetic moment in the continuum limit by studying the
	momentum dependence of the form factors.  We also compute the charm
	electromagnetic form factors within the same setup, which we
	find to be consistent with zero within the statistical
	precision of our data.
\end{abstract}

\maketitle

\section{Introduction}

Nucleon electromagnetic form factors serve as fundamental probes for
investigating the complex hadronic structure, as they reveal the
electric and magnetic distributions within nucleons. The strange quark
contribution, as the lightest non-valence quark in nucleons, is of
particular interest since it offers a distinct opportunity to examine
virtual particle dynamics in the nonperturbative regime of Quantum
Chromodynamics (QCD). Experimentally, strange electromagnetic form
factors are determined indirectly through parity-violating asymmetry
measurements, which arise from the interference between
electromagnetic and weak interactions resulting from electroweak
mixing~\cite{Kaplan:1988ku, Maas:2017snj}.
While several experiments, such as SAMPLE~\cite{SAMPLE:2003wwa,
Beise:2004py}, A4~\cite{A4:2004gdl,Maas:2004dh,Baunack:2009gy},
HAPPEX~\cite{HAPPEX:2005qax,HAPPEX:2005zgj,HAPPEX:2006oqy,HAPPEX:2011xlw}
and G0~\cite{G0:2005chy,G0:2009wvv}, do not exclude a zero value for
the strange magnetic moment, $\mu^s$, and electric and magnetic radii,
$\langle r^2_E\rangle^s$ and $\langle r^2_M\rangle^s$, respectively,
non-zero contributions to the nucleon electromagnetic form factors at
finite momentum transfer have been reported.  A first principles
lattice QCD calculation of the strange electromagnetic form factors is
thus valuable for providing {\it ab initio} input to these
experiments.

During the last decade, lattice QCD results on
the strange content of nucleons have been obtained typically using ensembles
with heavier than physical quark mass
parameters~\cite{Green:2015wqa,Djukanovic:2019jtp} while others
employing a single ensemble at physical quark masses, referred to here as physical
point~\cite{Sufian:2016pex,Alexandrou:2018zdf,Alexandrou:2019olr}.
These results consistently demonstrated that the lattice QCD framework can provide 
a robust  determination of the strange electromagnetic form
factors, thereby putting constraints on the determination of
the weak charge of the proton as explored by the $Q_{\rm weak}$
experiment~\cite{Qweak:2018tjf}.

In this work, we perform a calculation of the strange nucleon
electromagnetic form factors from lattice QCD using four ensembles of
clover-improved twisted mass fermions with two degenerate light,
strange, and charm quarks (\Nf{2}{1}{1}) with masses tuned to approximately their
physical values and similar physical volume. The four lattice
spacings are $a$=0.080~fm, 0.068~fm, 0.057~fm, and 0.049~fm, allowing us for the first time to  take the
continuum limit directly at the physical pion mass, without  the need for any chiral extrapolations.

We additionally compute the charm electromagnetic form factors of the
nucleon within the same setup, however we do not identify a clear
non-zero signal within the statistical precision of our data.
The charm contribution is expected to be significantly suppressed
owing to its heavier quark mass. However, a recent machine
learning-based study of the experimental data~\cite{Ball:2022qks}
points to a non-zero intrinsic charm parton distribution function
(PDF) in the proton, while a lattice QCD study dedicated to the charm
quark contribution using overlap valence fermions on Domain Wall gauge
configurations~\cite{Sufian:2020coz} obtains a small, non-zero
contribution.

The rest of this paper is structured as follows: 
The nucleon electromagnetic form factors, radii, and magnetic moments are defined in Sec.~\ref{sec:theory}; 
the lattice setup of our calculation, including statistics and details of the ensembles used, is described in Sec.~\ref{sec:lattice}; 
the method used to extract the form factors from the lattice QCD correlators is described in Sec.~\ref{sec:extraction}, 
along with method for identification of ground-state dominance and, the fits of the momentum transfer-dependence of the form factors. 
In Sec.~\ref{sec:fits}, we include the continuum extrapolation together with the results for the form factors for each gauge ensemble and fits to their momentum transfer-dependence. 
Sec.~\ref{sec:final_results} contains our final results and comparisons, whereas Sec.~\ref{sec:conclusions} summarizes and presents our conclusions. 
The full decomposition of the nucleon matrix elements in terms of the form factors is included in Appendix~\ref{sec:appendix_equations} for completeness, 
and tables with the numerical results for the electric and magnetic form factors are provided in Appendix~\ref{sec:appendix_results}.

\section{Decomposition of the nucleon matrix element}
\label{sec:theory}

We analyze the electromagnetic form factors of the nucleon in the SU(2)  isospin symmetric limit, i.e. neglecting isospin-breaking effects due to QED interactions and the u–d quark mass difference. This leads to proton and neutron being mass degenerate. The form factors in Minkowski space are given in terms of the matrix element of the electromagnetic current with nucleon states as follows
\begin{eqnarray}
	&& \langle N(p',s') \vert j_\mu \vert N(p,s) \rangle = \sqrt{\frac{m_N^2}{E_N(\vec{p}\,') E_N(\vec{p})}} \times   \label{eq:me} \\
	&& \bar{u}_N(p',s') \left[ \gamma_\mu F_1(q^2) + \frac{i \sigma_{\mu\nu} q^\nu}{2 m_N} F_2(q^2) \right] u_N(p,s)\,\nonumber,
\end{eqnarray}
where $N(p,s)$ is the nucleon with initial (final) momentum $p$ ($p'$) and spin $s$ ($s'$), with energy $E_N(\vec{p})$ ($E_N(\vec{p}\,') $) and mass $m_N$, $u_N$ is the nucleon spinor, $j_\mu$ is the vector current, and $q^2{\equiv}q_\mu q^\mu$ is the momentum transfer squared with $q_\mu{=}(p_\mu'-p_\mu)$. In Eq.~(\ref{eq:me}), the form factors of the nucleon are given in terms of the Dirac, $F_1$, and Pauli, $F_2$, form factors. These can be be linearly recombined to give the electric and magnetic Sachs form factors $G_E(q^2)$ and $G_M(q^2)$ via
\begin{gather}
	G_E(q^2) = F_1(q^2) + \frac{q^2}{4m_N^2} F_2(q^2)\,,\nonumber\\
	G_M(q^2) = F_1(q^2) + F_2(q^2)\,.\label{eq:sachs}
\end{gather}
The local vector current is given by,
\begin{equation}
	j_\mu = \sum_{\mathsf{f}=u,d,s,c} e_\mathsf{f} j^\mathsf{f}_\mu = \sum_{\mathsf{f}=u,d,s,c} e_\mathsf{f} \bar{\mathsf{\psi_f}} \gamma_\mu \mathsf{\psi_f}\,,\label{eq:current}
\end{equation}
where the sum over $\mathsf{f}$ runs over the up-, down-, strange- and
charm-quark flavors (u, d, s and c, respectively) and $e_\mathsf{f}$
is the electric charge of the quark with flavor $\mathsf{f}$. In this
work, we focus on $\mathsf{f}=s,c$. The light quark contributions to
the electromagnetic form factors have been presented in
Ref.~\cite{Alexandrou:2025vto} within the same setup and using the
three coarser of the four ensembles used here. Our final results are
provided without multiplying by the respective electric charges ($e_s$
and $e_c$).

At zero momentum transfer ($q^2=0$), the electric form factor yields
the contribution of the charge of the quark to the charge of the
hadron and the magnetic form factor the magnetic moment, i.e.
\begin{eqnarray}
	G_E^s(0) =& 0, \;\;  G_M^s(0) =& \mu^s.
\end{eqnarray}

The electric and magnetic radii are extracted by computing the slope of the corresponding Sachs form factor in the limit of $q^2\rightarrow 0$, namely
\begin{equation}
	\langle r_X^2 \rangle^{\mathsf{f}} = -6
	\frac{\partial G_X^\mathsf{f}(q^2)}{\partial q^2} \Big
	\vert_{q^2=0}\,,
	\label{eq:radius}
\end{equation}
where $X=E$ or $M$.

\section{Lattice setup}
\label{sec:lattice}

\subsection{Lattice correlation functions}
Within the framework of lattice QCD, the nucleon matrix elements can be accessed by appropriate combinations of two- and three-point correlation functions. These are defined in a Euclidean space-time, and we will thus use notation for Euclidean quantities from here on, including expressing the form factors in terms of $Q^2=-q^2$. We use the standard nucleon interpolating field,
\begin{equation}
	\chi_N(\vec{x},t)=\epsilon^{abc}u^a(x)[u^{b\intercal}(x)\mathcal{C}\gamma_5d^c(x)]\,,
\end{equation}
where $\mathcal{C}{=}\gamma_0 \gamma_2$ is the charge conjugation matrix.  The two-point function in momentum space is given by
\begin{align}
	C(\Gamma_0,\vec{p};t_s,t_0) = \sum_{\vec{x}_s} &
	e^{{-}i (\vec{x}_s{-}\vec{x}_0) \cdot \vec{p}}\label{eq:twop}                                                                                                          \\
	                                               & \mathrm{Tr} \left[ \Gamma_0 {\langle}\chi_N(t_s,\vec{x}_s) \bar{\chi}_N(t_0,\vec{x}_0) {\rangle} \right] \,,\nonumber
\end{align}
and the three-point function is given by
\begin{align}
	C_\mu(\Gamma_\nu,\vec{q},\vec{p}\,';t_s,\tins,t_0) {=} &
	\sum_{\vxins,\vec{x}_s}  e^{i (\vxins {-} \vec{x}_0)  \cdot \vec{q}}  e^{-i(\vec{x}_s {-} \vec{x}_0)\cdot \vec{p}\,'} \nonumber \\
	\mathrm{Tr} [ \Gamma_\nu \langle \chi_N(t_s,\vec{x}_s) & j_\mu(\tins,\vxins) \bar{\chi}_N(t_0,\vec{x}_0) \rangle].
	\label{eq:thrp}
\end{align}

The initial lattice site where the nucleon is created is denoted by  $x_0$ and referred to as the \textit{source}, the lattice site where the current couples to a quark  $\xins$ as the \textit{insertion}, and the site where the nucleon is annihilated  $x_s$ as the \textit{sink}. $\Gamma_\nu$ is a projector acting on Dirac indices, with $\Gamma_0 {=} \frac{1}{2}(1{+}\gamma_0)$ yielding the unpolarized and $\Gamma_k{=}\Gamma_0 i \gamma_5 \gamma_k$ the polarized matrix elements. Without loss of generality we will take $t_s$ and $\tins$ relative to the source time $t_0$ in what follows.

Inserting a complete set of states in Eq.~(\ref{eq:twop}) yields,
\begin{align}
	C(\vec{p},t_s) = & \sum_{n}c_n(\vec{p}) e^{-E_n(\vec{p}) t_s},\,\mathrm{where}\label{eq:twop_spec}                                                     \\
	c_n(\vec{p}) =   & \sum_{s}\mathrm{Tr}[ \Gamma_0 \langle \chi_N | n, \vec{p}, s \rangle \langle n, \vec{p}, s \vert \bar{{\chi}}_N  \rangle],\nonumber
\end{align}
with $| n, \vec{p}, s \rangle$ a QCD eigenstate with the quantum numbers of the nucleon, spin $s$, momentum $\vec{p}$, and energy $E_n(\vec{p})$. Similarly, inserting two complete sets of states in Eq.~(\ref{eq:thrp}) yields,
\begin{align}
	C_{\mu}(\Gamma_\nu,\vec{q},\vec{p}\,';t_s,\tins) =                         & \label{eq:thrp_spec}                                                                                              \\
	\sum_{n,m}  \mathcal{A}^{n,m}_{\mu}(\Gamma_\nu,\vec{q},\vec{p}\,')         &
	e^{-E_n(\vec{p}\,')(t_s-\tins)-E_m(\vec{q})\tins},\,\mathrm{where}\nonumber                                                                                                                    \\
	\mathcal{A}^{n,m}_{\mu}(\Gamma_\nu,\vec{q},\vec{p}\,') =                   & \nonumber                                                                                                         \\
	\sum_{s,s'}\mathrm{Tr}[\Gamma_\nu \langle \chi_N | n, \vec{p}\,',s'\rangle & \langle n, \vec{p}\,',s' | j_\mu | m, \vec{p}, s \rangle \langle m, \vec{p}, s | \bar{\chi}_N  \rangle].\nonumber
\end{align}
The desired nucleon matrix element is obtained for $m=n=0$ in Eq.~(\ref{eq:thrp_spec}) and after canceling the overlaps $\langle \chi_N | 0, \vec{p}\,',s'\rangle$ and $\langle 0, \vec{p}, s |\bar{\chi}_N \rangle$. One way to achieve this is by using an optimized ratio composed of two- and three-point functions, given by
\begin{align}
	R_{\mu}(\Gamma_{\nu},\vec{p},\vec{p}\,';t_s,\tins) = \frac{C_{\mu}(\Gamma_{\nu},\vec{p},\vec{p}\,';t_s,\tins\
	)}{C(\Gamma_0,\vec{p}\,';t_s)} \times \nonumber \\
	\sqrt{\frac{C(\Gamma_0,\vec{p};t_s-\tins) C(\Gamma_0,\vec{p}\,';\tins) C(\Gamma_0,\vec{p}\,';t_s)}{C(\Gamma_0,\vec{p}\,';t_s-\tins) C(\Gamma_0,\vec{p};\tins) C(\Gamma_0,\vec{p};t_s)}}.
	\label{eq:full_ratio}
\end{align}

In the limit of large time separations $(t_s-\tins) \gg$ and $\tins\gg$, the ratio in Eq.~(\ref{eq:full_ratio}) converges to the nucleon ground state matrix element, namely
\begin{gather}
	R_{\mu}(\Gamma_{\nu},\vec{p},\vec{p}\,';t_s,\tins) \xrightarrow[(t_s-\tins) \gg]{\tins\gg}\Pi_{\mu}(\Gamma_{\nu},\vec{p},\vec{p}\,').
\end{gather}

\subsection{Gauge ensembles and statistics}
\label{sec:ens}

We employ the twisted-mass fermion discretization
scheme~\cite{Frezzotti:2003ni,Frezzotti:2000nk}, which provides
automatic ${\cal O}(a)$-improvement~\cite{ETM:2010iwh}. In order to
obtain ensembles at approximately the physical isosymmetric pion mass
$m_\pi=135$~MeV~\cite{Alexandrou:2018egz,Finkenrath:2022eon}, tuning
of the bare light quark parameter, $\mu_l$ is performed. The heavy
quark parameters, $\mu_s$ and $\mu_c$ are tuned using the kaon mass
and an appropriately defined ratio between the kaon and D-meson
masses, as well as the D-meson mass, following the procedure of
Refs.~\cite{Finkenrath:2022eon,Alexandrou:2018egz}. The twisted mass
term in the action breaks  isospin symmetry at finite lattice
spacing, which is restored in the continuum limit. In order to alleviate
this isospin-breaking effect at finite lattice spacing, a clover term
is included in the action.  The parameters of the four ensembles
analyzed in this work can be found in Table~\ref{tab:ens}, where the
ensemble with the smallest lattice spacing, namely
\texttt{cE211.044.112}, is the new addition with respect to our previous
analysis reported in Ref.~\cite{Alexandrou:2025vto}. The lattice spacings and pion masses
are taken from Ref.~\cite{ExtendedTwistedMass:2022jpw} for the first
three ensembles and Ref.~\cite{ExtendedTwistedMass:2024nyi} for the
\texttt{cE211.044.112} ensemble. These values, obtained from the meson sector, are compatible with the
values determined from the nucleon mass in
Ref.~\cite{ExtendedTwistedMass:2021gbo}.

\begin{table}[h]
	\caption{Parameters of the four \Nf{2}{1}{1} ensembles
		analyzed in this work. From the leftmost to rightmost
		columns, we provide the name of the ensemble and its short
		acronym in parenthesis, the lattice volume, $\beta=6/g^2$
		with $g$ the bare coupling constant, the lattice spacing,
		the pion mass, and the value of $m_\pi L$. Lattice spacings
		and pion masses are taken from Refs.~\cite{ExtendedTwistedMass:2022jpw,ExtendedTwistedMass:2024nyi}.}  \label{tab:ens}
	\centering
	\begin{tabular}{cccccc}
		\hline\hline
		Ensemble                   & $\left(\frac{L}{a}\right)^3\times\left(\frac{T}{a}\right)$ & $\beta$ & \makecell[c]{$a$                   \\$[$fm$]$} & \makecell[c]{$m_\pi$\\ $[$MeV$]$} & $m_\pi L$ \\
		\hline
		\texttt{cB211.072.64}  (B) & $64^3 \times 128$                                          & 1.778   & 0.07957(13)      & 140.2(2) & 3.62 \\
		\texttt{cC211.060.80}  (C) & $80^3 \times 160$                                          & 1.836   & 0.06821(13)      & 136.7(2) & 3.78 \\
		\texttt{cD211.054.96}  (D) & $96^3 \times 192$                                          & 1.900   & 0.05692(12)      & 140.8(2) & 3.90 \\
		\texttt{cE211.044.112} (E) & $112^3 \times 224$                                         & 1.960   & 0.04892(11)      & 136.5(2) & 3.79 \\
		\hline
	\end{tabular}
\end{table}

To increase the overlap of the interpolating field with the nucleon
ground state, so that excited states are suppressed at earlier time
separations, we use Gaussian smeared point
sources~\cite{Gusken:1989qx,Alexandrou:1992ti},
\begin{equation}
	\psi^\mathrm{sm}(\vec{x}, t) = \sum_{\vec{y}} [\mathbf{1} +
		a_G H(\vec{x}, \vec{y}; U(t))]^{N_G} \psi(\vec{y},t),
\end{equation}
where the hopping matrix is given by
\begin{eqnarray}
	H(\vec{x},\vec{y};U(t)) = \sum_{i=1}^3 \left[ U_i(x) \delta_{x,y-\hat{i}} + U_i^\dag(x-\hat{i}) \delta_{x,y+\hat{i}}  \right].
\end{eqnarray}
The parameters $a_G$ and $N_G$ are
tuned~\cite{Alexandrou:2018sjm,Alexandrou:2019ali} in order to
approximately give a root mean square (r.m.s) radius for the nucleon
of $\approx 0.5$~fm. For the links entering the hopping matrix, we apply APE
smearing~\cite{APE:1987ehd} to reduce statistical errors due to
ultraviolet fluctuations. The smearing parameters are tuned separately
for each value of the lattice spacing and are given in
Table~\ref{tab:smearing}.

\begin{table}
	\caption{Number of Gaussian smearing iterations ($N_G$) and Gaussian
		smearing parameter $a_G$ used for each ensemble. We also provide
		the number of APE-smearing iterations $n_{\rm APE}$ and parameter
		$\alpha_{\rm APE}$ applied to the links that enter the Gaussian
		smearing hopping matrix.}\label{tab:smearing}
	\begin{tabular}{cccccc}
		\hline\hline
		Ensemble                   & $N_G$ & $a_G$ & $n_{\rm APE}$ & $\alpha_{\rm APE}$ \\
		\hline
		\texttt{cB211.072.64} (B)  & 125   & 0.2   & 50            & 0.5                \\
		\texttt{cC211.060.80} (C)  & 140   & 1.0   & 60            & 0.5                \\
		\texttt{cD211.054.96} (D)  & 200   & 1.0   & 60            & 0.5                \\
		\texttt{cE211.044.112} (E) & 250   & 1.0   & 60            & 0.5                \\
		\hline
	\end{tabular}
\end{table}

\begin{figure}
	\includegraphics[width=0.7\linewidth]{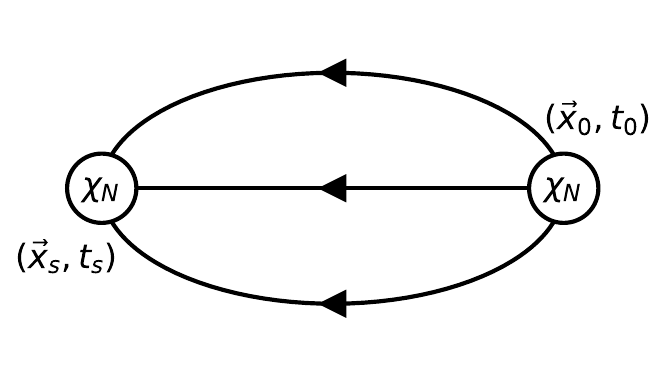}
	\includegraphics[width=0.7\linewidth]{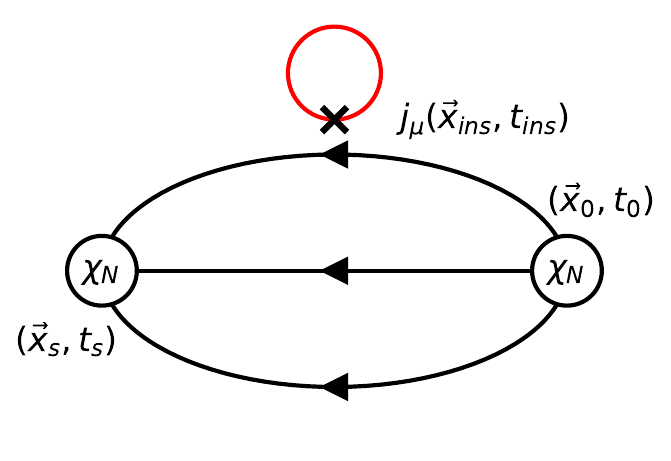}
	\caption{Nucleon two-point function (top) and disconnected nucleon three-point function (bottom). The red lines denote all-to-all
		quark propagators.}\label{fig:diagrams}
\end{figure}
Carrying out the quark field contractions for the three-point function
in Eq.~(\ref{eq:thrp}) gives rise to quark disconnected contributions
for the case of strange and charm quarks in the current, as shown in
the bottom diagram of Fig.~\ref{fig:diagrams}.  The disconnected
three-point functions involve correlating the two-point function of
Eq.~(\ref{eq:twop}) with the quark loop ($L_\mathsf{f}$) with the
electromagnetic current insertion,
\begin{equation}
	L_\mathsf{f}(\tins,\vec{q}) = \sum_{\vec{x}_\mathrm{ins}} \mathrm{Tr}
	\left[ D_\mathsf{f}^{-1}(x_\mathrm{ins};x_\mathrm{ins}) \gamma_\mu \right].
	e^{i \vec{q} \cdot \vec{x}_\mathrm{ins}},\label{eq:looptrace}
\end{equation}
for each quark flavor $\mathsf{f}$. The strange and charm quark loops
are computed stochastically, via the {generalized one-end
		trick}~\cite{McNeile:2006bz}, using volume stochastic sources, as also
used in our previous
studies~\cite{Alexandrou:2013wca,Alexandrou:2017hac,Alexandrou:2017qyt,Alexandrou:2017oeh,Alexandrou:2018sjm}.

Note that the two-point correlation function is obtained for any sink
momentum $\vec{p}\,'$, which allows us to correlate with the loop to
obtain the three-point correlation function for multiple sink momenta
at no additional computational cost. Also, since the two-point
correlation function is obtained for all $t_s$, we can obtain
the three-point function for any $\tins$ and all insertion and sink time
combinations. In practice however, we are limited by statistical
uncertainties, which grow with $t_s$ and $\vec{p}\,'$.

\begin{table}
	\caption{Parameters for the computation of the disconnected
		three-point function. For the four ensembles (indicated in
		the first column), we give the number of configurations
		($n_\mathrm{conf}$, second column) and the number of source
		positions for the two-point functions ($n_\mathrm{src}$,
		third column). For each of the strange and charm
		($\mathsf{f}$, third column) we provide the number of
		vectors that are inverted per configuration
		($n_\mathrm{vec}$, fourth column) resulting from the product
		of vectors needed for color and spin dilution
		($n_\mathrm{dil}$), hierarchical probing ($n_\mathrm{had}$),
		and the number of stochastic vectors used
		($n_\mathrm{stoch}$).}\label{tab:stoch}
	\begin{tabular}{cccc r@{$\times$}c@{$\times$}l }
		\hline\hline
		\multirow{2}{*}{Ensemble}               & \multirow{2}{*}{$n_\mathrm{conf}$} & \multirow{2}{*}{$n_\mathrm{src}$} & \multirow{2}{*}{$\mathsf{f}$} & \multicolumn{3}{c}{$n_\mathrm{vec}$=}                                         \\
		                                        &                                    &                                   &                               & $n_\mathrm{dil}$                      & $n_\mathrm{had}$ & $n_\mathrm{stoch}$ \\
		\hline
		\multirow{2}{*}{\texttt{cB211.072.64}}  & \multirow{2}{*}{749}               & \multirow{2}{*}{349}              & s                             & 12                                    & 512              & 1                  \\
		                                        &                                    &                                   & c                             & 12                                    & 32               & 12                 \\
		\hline
		\multirow{2}{*}{\texttt{cC211.060.80}}  & \multirow{2}{*}{399}               & \multirow{2}{*}{650}              & s                             & 12                                    & 512              & 4                  \\
		                                        &                                    &                                   & c                             & 12                                    & 512              & 1                  \\
		\hline
		\multirow{2}{*}{\texttt{cD211.054.96}}  & \multirow{2}{*}{493}               & \multirow{2}{*}{368}              & s                             & 12                                    & 512              & 4                  \\
		                                        &                                    &                                   & c                             & 12                                    & 512              & 1                  \\
		\hline
		\multirow{2}{*}{\texttt{cE211.044.112}} & \multirow{2}{*}{501}               & \multirow{2}{*}{339}              & s                             & 12                                    & 512              & 2                  \\
		                                        &                                    &                                   & c                             & 12                                    & 512              & 1                  \\
		\hline
	\end{tabular}

\end{table}

The parameters used for the computation of the disconnected
three-point function are summarized in Table~\ref{tab:stoch} for both
strange and charm. For all four ensembles we fully dilute in color and
spin ($n_\mathrm{dil}$=12). Furthermore, we use hierarchical
probing~\cite{Stathopoulos:2013aci} to distance eight in the
4-dimensional volume, which results in 512 Hadamard vectors
($n_\mathrm{had}$=512). An exception is the \texttt{cB211.072.64}
ensemble where we used $n_\mathrm{had}=32$ and 12 stochastic vectors
for the case of the charm loops. Hierarchical probing suppresses
stochastic noise from the off-diagonal spatial indices in the trace of
Eq.~(\ref{eq:looptrace}). We therefore use the local vector current,
rather than the conserved vector current, namely
$\mathsf{\bar{\psi_f}}\gamma_\mu\mathsf{\psi_f}$, which only couples
the diagonal components of the spatial indices of the insertion. This
requires us to renormalize the matrix element, as will be explained in
Sec.~\ref{subsec:renorm}.

We emphasize the high statistics used for this calculation, which can
be seen in Table~\ref{tab:stoch}. Namely, for each ensemble, we
compute $\mathcal{O}(10^5)$ two-point functions and use on each
configuration $\mathcal{O}(10^3)$ vectors for the loops.

\subsection{Renormalization}
\label{subsec:renorm}

\begin{figure}[ht!]
	\centering
	\includegraphics[width=\columnwidth]{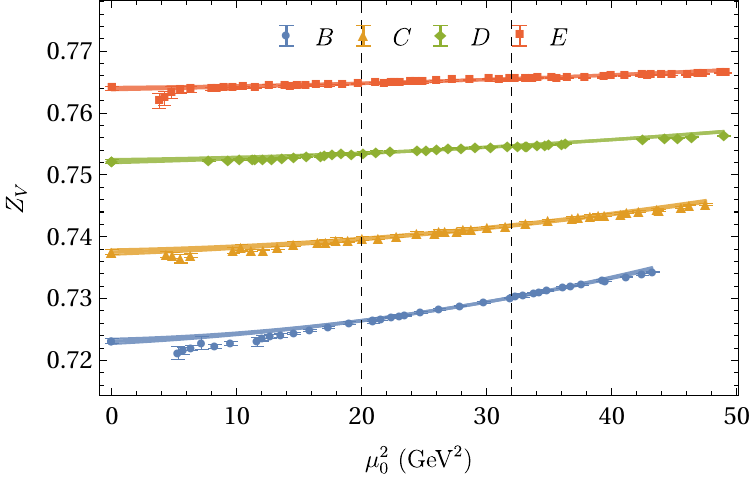}
	\caption{$Z_{\rm V}$ as a function of $\mu_0^2$ for the four lattice-spacing ensembles, namely B, C, D and E. A joint fit of the form $f (a^2 \mu_0^2) = c_0 (a) + c_1 a^2 \mu_0^2 + c_2 (a^2 \mu_0^2)^2$ is employed in the range $\mu_0^2 \in [20,32]$ GeV$^2$. The extrapolated values $c_0 (a)$ for each lattice spacing are given at $\mu_0^2 = 0$.}
	\label{fig:Zfactors}
\end{figure}

\begin{figure}[ht!]
	\centering
	\includegraphics[width=\columnwidth]{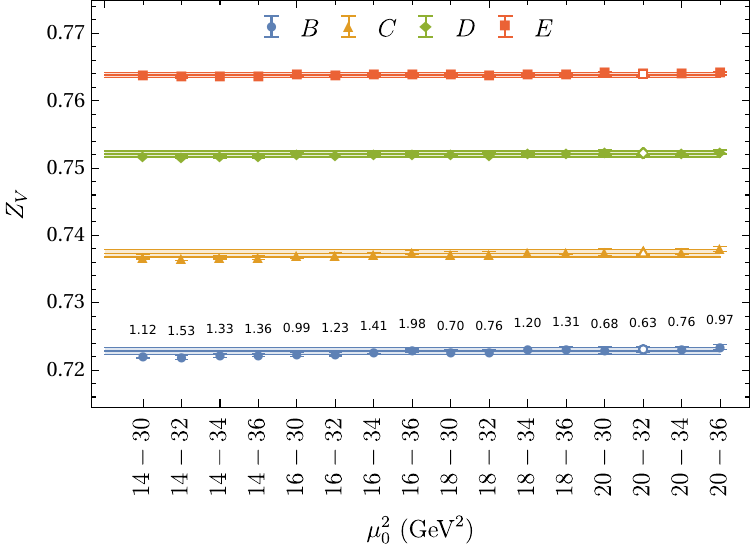}
	\caption{Extrapolated values of $Z_{\rm V}$ at $a^2 \mu_0^2 = 0$ from momentum fits over multiple fit ranges together with the model-averaged values (bands). The reduced $\chi^2$ of each fit is given by the label above the corresponding data point. The open symbols denotes the fit with the highest weight based on the Akaike Information Criterion.}
	\label{fig:AIC}
\end{figure}

The matrix elements of the local vector current are renormalized nonperturbatively using the RI$'$/MOM scheme~\cite{Martinelli:1994ty}. Since the vector current is scheme and scale independent, no perturbative conversion to a reference scheme or evolution to a reference scale is required. Furthermore, as shown in our calculation of Ref.~\cite{Alexandrou:2025vto}, the renormalization factors of flavor-singlet and nonsinglet vector operators coincide. Therefore, disconnected diagrams can be safely neglected in the renormalization analysis.
Following Ref.~\cite{Alexandrou:2025vto}, we extract the renormalization factor of the vector current, denoted by $Z_V$, using the condition~\cite{ETMC-ren:2025}:
\begin{equation}
	{(Z_q^{{\rm RI}'})}^{-1} Z_V \frac{1}{36} \sum_\mu {\rm Tr} \left[\Lambda_{V_\mu} (p)(\gamma_\mu - \frac{\slashed{\tilde{p}}}{4 \tilde{p}_\mu})\right] \Bigg |_{p^2 = \mu_0^2} = 1,
\end{equation}
where $\tilde{p}_\mu \equiv \sin(a p_\mu)\neq 0$ and $\mu_0$ denotes the RI$'$/MOM scale. $\Lambda_{V_\mu}(p)$ is the amputated vertex function of the vector operator with external off-shell quark states, and $Z_q^{{\rm RI}'}$ is the renormalization factor of the quark field obtained from the quark propagator~\cite{ExtendedTwistedMass:2021gbo}.
The procedure follows Ref.~\cite{Alexandrou:2025vto} and is summarized below. Vertex functions and quark propagators are computed in Landau gauge using momentum sources~\cite{Gockeler:1998ye} on 30 configurations. We use ensembles generated specifically for the renormalization program (see Ref.~\cite{Alexandrou:2024ozj}), which contain four mass-degenerate quarks (\Nf{4}) and are simulated at the same $\beta$ values as the four physical-point ensembles used for the matrix-element analysis. The ensembles correspond to different twisted-mass parameters $\mu_{\rm sea}$, allowing an extrapolation to the chiral limit. The dependence on $\mu_{\rm sea}$ is found to be mild, consistent with previous studies of our group (see, e.g., Ref.~\cite{Alexandrou:2015sea}). This dependence is removed through a linear fit in $\mu_{\rm sea}$.
To suppress rotational $O(4)$ breaking effects, we employ spatially isotropic momenta satisfying the ``democratic'' criterion: $\sum_{\mu} p_\mu^4/(\sum_\mu p_\mu^2)^2 < 0.3$. In addition, we improve the nonperturbative results by subtracting one-loop discretization effects computed in lattice perturbation theory to all orders in the lattice spacing~\cite{Alexandrou:2015sea}.
Residual lattice artifacts depending on the renormalization scale $\mu_0$ are removed through polynomial fits in $a^2 \mu_0^2$. Figure~\ref{fig:Zfactors} shows a representative fit in the range $20$--$32~{\rm GeV}^2$. Compared to the analysis of Ref.~\cite{Alexandrou:2025vto}, we additionally determine the renormalization factor for the finer E lattice spacing. Moreover, in the present work we perform a joint fit in $a^2\mu_0^2$ using all four lattice spacings, rather than fitting each ensemble separately. To accommodate the combined dataset, higher-order terms in $a^2\mu_0^2$ are included in the fit ansatz. Specifically, we consider $f (a^2 \mu_0^2) = c_0 (a) + c_1 a^2 \mu_0^2 + c_2 (a^2 \mu_0^2)^2$. These simultaneous fits provide an improved description of the data.
We perform fits over several ranges within $14$--$36~{\rm GeV}^2$. The extrapolated values at $\mu_0=0$ from all fits are combined using model averaging (see Sec.~\ref{sec:MA}). Figure~\ref{fig:AIC} displays the results from all fits together with the model-averaged value. The final values of $Z_V$ are reported in Table~\ref{tab:disc_range}. Note that a comprehensive analysis based on a total of five lattice spacings, will be presented in a forthcoming publication by ETMC~\cite{ETMC-ren:2025}.

\section{Extraction of nucleon matrix elements}
\label{sec:extraction}

In order to obtain the bare form factors at each value of the momentum
transfer squared, $Q^2$, we need to combine $\Gamma_\nu$ and $\mu$
depending on the momenta $\vec{p}\,'$ and $\vec{q}$ in the ground
state matrix element, $\Pi_\mu(\Gamma_\nu;\vec{p}\,', \vec{q})$ in
order to separate $G_E(Q^2)$ and $G_M(Q^2)$.  The sink momentum in the
two-point correlation function can be selected independently of the
momentum flowing through the disconnected loop at no additional
computational cost. We thus take sink momenta
$\vec{p}\,'=\frac{2\pi}{L}\vec{k}$ with $\vec{k}^2=1$ and $2$ in
addition to the case $\vec{p}\,'=0$. With these additional momenta,
the expressions yielding $G_E$ and $G_M$ cannot be disentangled (see
for example Appendix~\ref{sec:appendix_equations}), i.e. each
combination of $\Gamma_\nu$, $\mu$, and sink and insertion momenta in
$\Pi_\mu(\Gamma_\nu, \vec{p}\,', \vec{q})$ relates to a linear
combination of $G_E(Q^2)$ and $G_M(Q^2)$. In the resulting system of
equations, the number of independent constraints exceeds the number of
unknown parameters to be determined. Consequently, a direct inversion
is not feasible, and we instead employ a Singular Value Decomposition
(SVD) approach to obtain a stable and well-defined solution for the
electromagnetic form factors. The SVD framework allows us to
systematically handle correlations and to mitigate the effects of
numerical instabilities that typically arise in overconstrained
systems.  Proceeding similarly to our approach in
Ref.~\cite{Alexandrou:2019ali, Alexandrou:2025vto}, including for the
propagation of statistical errors, we extract the electric and
magnetic form factors separately as a function of the momentum
transfer squared.

\begin{table}
	\centering
	\caption{Values of the source-sink separations of the
		three-point function ($t_s$) and the minimum insertion used
		in plateau fits ($\tins^{\rm min}$) with maximum insertion,
		$\tins^{\rm max} = t_s - \tins^{\rm min}$. We also list the
		renormalization constants of the singlet current, $Z_V$,
		used in our analysis and the nucleon mass, $m_{N}$ used in
		the SVD.}
	\label{tab:disc_range}
	\begin{tabular}{cccccc}
		\hline\hline Ensemble & $t_s/a$                & $\tins^{\rm min}/a$ & $Z_V$     & $m_{N}$ [GeV] \\
		\hline
		B                     & 10, 12, 14, 16         & 3                   & 0.7228(5) & 0.9372(50)    \\
		C                     & 12, 14, 16, 18         & 3                   & 0.7373(5) & 0.9420(51)    \\
		D                     & 14, 16, 18, 20, 22     & 4                   & 0.7521(4) & 0.9431(38)    \\
		E                     & 16, 18, 20, 22, 24, 26 & 5                   & 0.7638(4) & 0.9391(38)    \\
		\hline
	\end{tabular}
\end{table}

\begin{figure}[h]
	\centering
	\includegraphics[width=\linewidth]{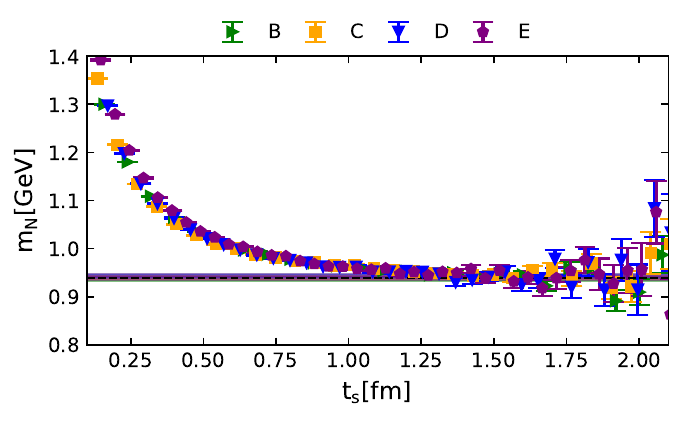}
	\caption{Nucleon effective mass vs the source-sink time
		separation, $t_s$ for the ensembles analysed in this
		work. As indicated in the header, the right-pointing green
		triangles, orange squares, downward pointing blue triangles,
		and the purple pentagons denote the results for the coarsest
		to the finest lattice spacing used in this work. The
		different color bands are the result of the effective mass
		of the nucleon using a three-state fit, whereas the black
		dashed line corresponds to the extracted mass of the nucleon
		$m_{N} = 0.938 \;\rm GeV$.}
	\label{fig:meff}
\end{figure}

\begin{figure}[h]
	\centering
	\includegraphics[width=\linewidth]{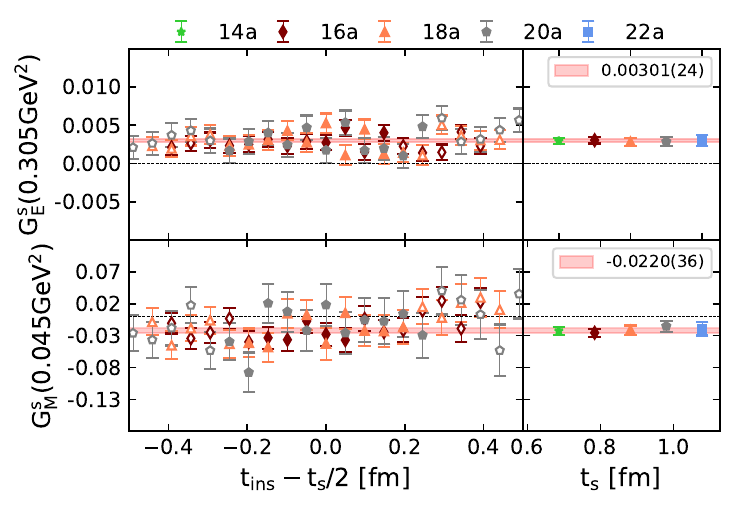}
	\caption{Results on  $G_E^s(Q^2=0.305 \rm \;GeV^2)$ (top)
		and $G_M^s(Q^2=0.045 \rm \;GeV^2)$ (bottom) for the
		\texttt{cE211.044.112} ensemble. The left column shows
		source-sink time separations $t_s = \rm 16a, 18a$ and $20\rm a$
		as indicated in the header, for all insertion times $\tins$,
		where the filled points denote those included in the plateau
		fits. The right column shows the result using plateau fits
		for $t_s = \rm 14a, 16a, 18a, 20a$ and $22\rm a$. The red band denotes the weighted average of results for all $t_s$ as listed in Table~\ref{tab:disc_range}, taken as final value.}
	\label{fig:E_disc}
\end{figure}

To summarize, for a given $Q^2$, with $N$ equations, we have
\begin{equation}
	\Pi_\mu(\Gamma_{\nu};\vec{p}\,', \vec{p}) = \mathcal{G}^{\mu\nu}(\vec{p}\,', \vec{p}) \; F(Q^2),
	\label{eq:ratio_matrix_eqn}
\end{equation}
where the matrix $\mathcal{G}$ is an $N\times2$ array of kinematic factors and $F(Q^2)^\top = [G_E(Q^2), G_M(Q^2)]$. Taking an SVD of the coefficient matrix $\mathcal{G}$,
\begin{equation}
	\mathcal{G} = U\Sigma V
	\label{eq:SVD}
\end{equation}
we can then extract the form factors by using
\begin{equation}
	F(Q^2) = V^{\dagger}\Sigma^{-1}U^{\dagger}\Pi(\vec{p},\vec{p}').
	\label{eq:SVD_extraction}
\end{equation}
Specifically, we proceed as follows;
\begin{itemize}
	\item We obtain the SVD of $\mathcal{G}$ for a given
	      $(\vec{p}^2, \vec{p}'^2)$. Note that multiple combinations
	      of $(\vec{p}^2, \vec{p}'^2)$ may yield the same $Q^2$. The
	      nucleon mass that appears in the coefficient matrix,
	      $\mathcal{G}$, is set from three-state fits to the two-point
	      functions. The nucleon effective mass for all ensembles used
	      in this work is shown in Fig.~\ref{fig:meff} and
	      the resulting nucleon mass is listed in Table~\ref{tab:disc_range}.

	\item We perform {plateau fits}, defined as a constant
	      fit in $\tins$ to $R_\mu(\Gamma_\nu, \vec{p}, \vec{p}';
		      t_s,\tins)$, which is the ratio of Eq.~(\ref{eq:full_ratio})
	      for a given source-sink time separation ($t_s$) and momentum
	      combinations $(\vec{p}^2, \vec{p}'^2)$ to obtain
	      $\Pi_\mu(\Gamma_\nu, \vec{p}, \vec{p}'; t_s)$.

	\item We compute $\tilde{\Pi}_{\mu\nu}(\vec{p},\vec{p}';t_s)
		      \equiv U^{\dagger}\Pi_\mu(\Gamma_\nu, \vec{p}, \vec{p}';
		      t_s)$, where only the first two rows of $\tilde{\Pi}$
	      contribute to the form factors.

	\item The results are then multiplied by
	      $V^{\dagger}\Sigma^{-1}$ yielding the two form factors. We
	      average over all $(\vec{p},\vec{p}')$ combinations
	      contributing to the same $Q^2$.
\end{itemize}
The process is carried out within a jackknife procedure, yielding jackknife
errors for the form factors at each $Q^2$.

We demonstrate this fitting procedure in Fig.~\ref{fig:E_disc}, where we
show the ratio formed for the strange form factor for the
\texttt{cE211.044.112} for $Q^2$ values where the form factors are expected to be large in magnitude, namely  $Q^2 = 0.305$ GeV$^2$ for $G_E(Q^2)$ and
$Q^2 = 0.045$ GeV$^2$ for $G_M(Q^2)$.
Convergence in $t_s$ is demonstrated by examining the value resulting
from plateau fits to the ratio with increasing source-sink time
separation $t_s$, as shown in Fig.~\ref{fig:E_disc}. We perform the
plateau fits to individual values
of $t_s$, where we start fitting from the minimum
insertion time, $\tins^{\rm min}$, up to the maximum insertion time,
$\tins^{\rm max}$ (see Table~\ref{tab:disc_range}).
We observe results are stable as we
increase $\tins^{\rm min}$ indicating that excited states are
suppressed. Therefore, we perform weighted average of the result of
the fits to $t_s\in[0.8, 1.28]\;\rm fm$ to extract our final result,
given in lattice units in Table~\ref{tab:disc_range} for each
ensemble.

\begin{figure*}
	\centering
	\includegraphics[width=\textwidth]{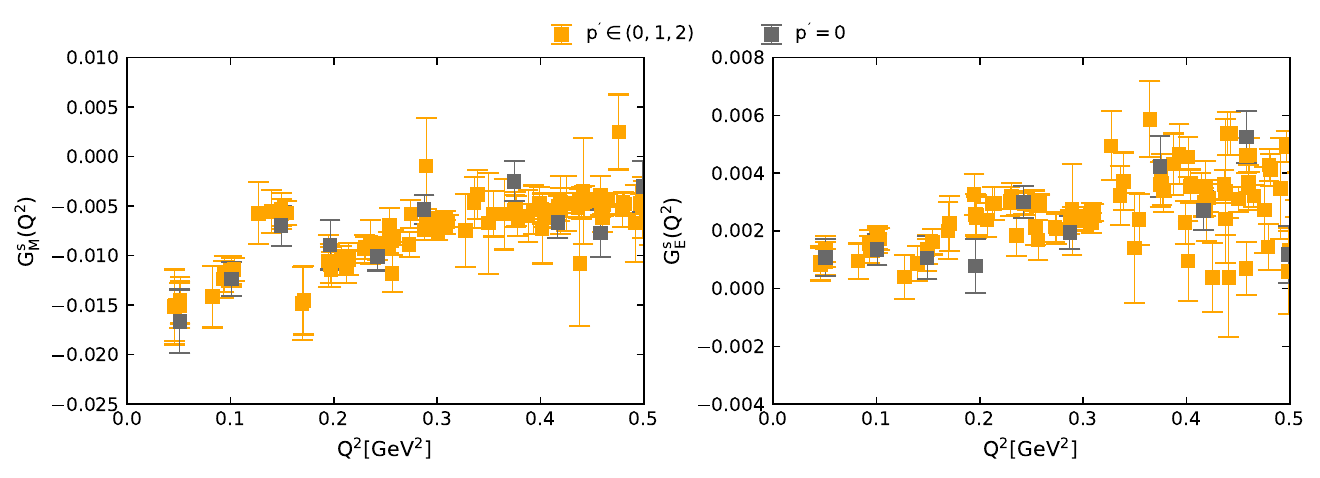}
	\caption{$G^s_M(Q^2)$ (left) and $G^s_E(Q^2)$ (right) form
		factors as a function of $Q^2$ for the ensemble
		\texttt{cC211.060.80} obtained by combining finite sink momentum values
		$p'^2\in(0,1,2)$ (orange squares) and when setting sink
		momentum to zero (gray squares).}
	\label{fig:pp0_pp2_compare}
\end{figure*}

When including non-zero sink momenta, the number of $Q^2$ values
within $Q^2\in [0,1)$~GeV$^2$ is 226 for \texttt{cB211.072.64}, 263
for \texttt{cC211.060.80}, 261 for \texttt{cD211.054.96}, and 268
for \texttt{cE211.044.112}. For presentation of the form factors and
further analysis, we will follow the approach adopted in our
previous works~\cite{Alexandrou:2019olr,Alexandrou:2025vto}, where
we average the form factors over small intervals of $Q^2$ i.e. we
average over intervals of $\delta Q^2$=0.04~GeV$^2$. For
completeness, we show the data before this binning process in
Fig.~\ref{fig:pp0_pp2_compare}, namely we show the data obtained for
ensemble \texttt{cC211.060.80} when we only use $\vec{p}'^2=0$ and
when using $p'^2\in(0,1,2)$ for $Q^2\in[0,0.5]$~GeV$^2$.

\section{Analysis of extracted form factors}
\label{sec:fits}

We investigate
the $Q^2$- and $a$-dependence of the form factors in two ways, namely
(i) via a \textit{``two-step approach''}, where the $Q^2$-dependence
of each ensemble is fitted separately, and then in a second step we
take the continuum limit as an extrapolation of the fit parameters in $a^2$, and (ii) via a
\textit{``one-step approach''}, where we augment the parameters of the
$Q^2$ dependence to include $a^2$-dependent terms, thereby obtaining
the $Q^2$ dependence and continuum limit within a single fit.  To combine the statistics of
the four ensembles in these fits, we compute the covariance matrix
constructed using the jackknife results per ensemble extracted as described in the
previous section.

The various parameterizations used for the $Q^2$-dependence are given in
the following subsections.
\subsection{Dipole form}
For the magnetic form factor, we use the standard dipole form given by
\begin{equation}
	G(Q^2) = \frac{g}{\left[1+\frac{Q^2\langle r'^2 \rangle}{12}\right]^2},
	\label{eq:dipole}
\end{equation}
with $g$ and $\langle r'^2 \rangle$ treated as fit parameters where the former
gives direct access to the magnetic moment and the radius is given by $\langle r^2 \rangle = g\langle r'^2 \rangle$.

For the one-step process, the $a$-dependence is incorporated within the parametrization of the fit parameters, namely
\begin{equation}
	g(a^2)=g_0+a^2g_2\,\,\text{and}\,\, \langle
	r'^2(a^2)\rangle=\langle r'^2\rangle_0+a^2\langle r'^2\rangle_2
	\nonumber
	\label{eq:dipole_r_mu}
\end{equation}
where, $g_0$ is the magnetic moment in
the continuum limit and the continuum limit radius is given by and $\langle r^2\rangle_0 = g_0\langle r'^2\rangle_0$. Parameters $g_2$ and $\langle r'^2\rangle_2$ capture
cutoff effects starting at $a^2$ as we have $O(a)$ improvement.  We
then substitute  in Eq.~(\ref{eq:dipole}) to obtain,
\begin{equation}
	G(Q^2,a^2) = \frac{g(a^2)}{\left[1+\frac{Q^2}{12}\langle r'^2(a^2)\rangle
			\right]^2}.
	\label{eq:dipole_a}
\end{equation}
In the two-step approach, the $a$-dependence of the fitted parameters is the same but the continuum limit is taken in a second step after fitting the $Q^2$ of each ensemble. This applies to all parametrizations used.

\subsection{Galster-like parameterization}
Since the strange electric form factor vanishes at $Q^2=0$ due to the
zero net strangeness of the nucleon, we use the Galster-like
parameterization~\cite{Galster:1971kv}, which is a modified version of
the dipole form such that $G(Q^2=0)=0$,
\begin{equation}
	G(Q^2) = \frac{Q^2 A}{4 m_N^2 + Q^2 B} \frac{1}{\left(1+\frac{Q^2}{0.71 {\rm GeV}^2}\right)^2},
	\label{eq:Galster-like}
\end{equation}
where $A$ and $B$ are the fit parameters.
In this case, the radius is given by
\begin{equation}
	\langle r^2 \rangle = - \frac{3A}{2 m_N^2}.
	\label{eq:Galster_radii}
\end{equation}

Similarly to the dipole form, we parametrised the $a$-dependence to
\begin{gather}
	A(a^2)=A_0 + a^2A_2 \nonumber\\
	B(a^2)=B_0 + a^2B_2
	\label{eq:galster_r_mu}
\end{gather}
such that $A_0$ and $B_0$ are the continuum limit
parameters. Substituting the above in Eq.~(\ref{eq:Galster-like}) we
have,
\begin{gather}
	G(Q^2, a^2) = \frac{Q^2 A(a^2)}{4 m_N^2 + Q^2 B(a^2)}\frac{1}{\left(1+\frac{Q^2}{0.71 {\rm GeV}^2}\right)^{2}}.
	\label{eq:Galster-like_a}
\end{gather}

For two-step Galster-like procedure, the nucleon mass $m_N$ is taken for each ensemble from Table~\ref{tab:disc_range}, whereas in the one-step approach it is fixed to the physical value $m_N = 0.938 \;\rm GeV$.

\subsection{$z$-expansion}
For both electric and magnetic form factors we also use the $z$-expansion parameterization given by,
\begin{equation}
	G(Q^2) = \sum_{k=0}^{k_{\rm max}} c_k z^k(Q^2),
	\label{eq:zexp}
\end{equation}
where $c_k$ are the fit parameters and
\begin{equation}
	z(Q^2) = \frac{\sqrt{t_{\rm cut} + Q^2} - \sqrt{t_{\rm cut}+t_0} }{ \sqrt{t_{\rm cut} + Q^2} + \sqrt{t_{\rm cut}+t_0} }
	\label{eq:zQ2}
\end{equation}
with $t_{\rm cut}$ being the particle production threshold chosen to
be $t_{\rm cut}=4m_K^2$ for the strange form factors considered here,
and $t_0$ an arbitrary choice in $(-t_{\rm cut}, \infty)$. We set
$t_0=0$, which simplifies the relation of the magnetic moment  and radii to the
fit parameters, namely
\begin{equation}
	g = c_0\quad\text{and}\quad \langle r^2\rangle = -\frac{3c_1}{2t_{\rm
				cut}}\quad\text{with}\quad t_0=0.
\end{equation}
If using a one-step approach, the $a^2$-dependence is again incorporated in the parameters i.e.,
\begin{equation}
	c_k(a^2)=c_{k,0}+a^2c_{k,2}
\end{equation}
which modifies the $z$-expansion as
\begin{equation}
	G(Q^2,a^2) = \sum_{k=0}^{k_{\rm max}} c_k(a^2) z^k(Q^2).
	\label{eq:zexp_a2}
\end{equation}
As pointed out in Ref.~\cite{Lee_2015}, in order to enforce smooth
convergence of the form factor to zero at $Q^2\rightarrow \infty$, we
have the following constraints,
\begin{equation}
	\sum_{k=0}^{k_{\rm max}} c_k \frac{d^nz^k}{dz^n}\Bigg|_{z=1} =
	0\quad\text{with}\quad n=0,1,2,3.
	\label{eq:inf_Q2_convergence}
\end{equation}
Since the statistical errors of our data do not permit arbitrarily
large values of $k_{\rm max}$, it is sufficient for us to ensure this
condition is met for $n=0$ and 1, leading to two constraints, namely
$\sum_{k=0}^{k_{\rm max}} c_{k,0} = 0$ and $\sum_{k=1}^{k_{\rm max}}
	kc_{k,0} = 0$. This implies that for a given $k_{\rm max}$, instead of
$k_{\rm max}+1$ fit parameters we have $k_{\rm max}-1$, with the last
two being fixed by these conditions. In order to obtain a stable fit,
we use Gaussian priors for our fit parameters following the approach
introduced in our previous work~\cite{Alexandrou:2023qbg} for the
axial form factors,  as well as in other similar analyses, see e.g.
Refs.~\cite{Meyer:2016oeg,Hill:2010yb}. Namely, we use
\begin{equation}\label{eq:zexp_priors}
	c_{k,0}\sim 0(w/k),\quad c_{k,2}\sim 0(20w/k)\quad\text{for}\quad k\ge3,
\end{equation}
where, $w\le9$ is a hyper-parameter controlling the width of the
priors and which we vary independently of $k_{\max}$.

\subsection{Model Averaging}
\label{sec:MA}
In order to combine the results from the different fit Ans\"atze and the
various momentum cuts used in the fits, we employ a model averaging
approach. Namely, we use the $\chi^2$ value and the degrees of freedom
of the fit to evaluate its probability via the Akaike Information
Criterion (AIC)~\cite{Jay:2020jkz,Neil:2022joj} and to obtain combined
statistical and systematic errors.  To summarize, for each combination
of fit Ansatz and fit ranges, labeled as ``model $i$'', we associate a
weight, $w_{i}$, defined as,
\begin{equation}\label{eq:weight}
	\log(w_{i}) = -\frac{\chi_{i}^2}{2} + N^{\rm dof}_{i},
\end{equation}
where $N^{\rm dof}_{i} = N^{\rm data}_{i} - N^{\rm params}_{i}$ is the
number of degrees of freedom, given as the difference between the
number of data, $N^{\rm data}_{i}$, and the number of parameters,
$N^{\rm params}_{i}$, used in the corresponding fit.
$\chi^2_{i}$ is defined as
\begin{equation}
	\chi^2_{i} = {(\vec{r_i})}^{\,\top} (C)_i^{-1}
	\vec{r}_{i}\quad\text{with}\quad
	\vec{r}_{i}=\vec{y}_{i}-f(\vec{x}_{i}),
\end{equation}
where, for each model $i$, $C_i$ is the covariance matrix of the
data determined via jackknife resampling,
$\vec{y}_{i}$ is the jackknife mean for model $i$, and $\vec{r}_{i}$
the residual, computed using the selected model
$f(\vec{x}_{i})$. The probability associated with each model is given by,
\begin{equation}
	p_{i} = \frac{w_{i}}{Z}\quad\text{with}\quad Z=\sum_i{w_{i}}
\end{equation}
and the model-averaged central value ($\bar{\mathcal{O}}$)
and total error ($\sigma_{\bar{\mathcal{O}}}$) of our
observable is given by:
\begin{equation}\label{eq:model_average}
	\bar{\mathcal{O}} = \sum_i p_{i} \mathcal{\bar{O}}_i\,\,\,\mathrm{and}\,\,\,\sigma^2_{\bar{\mathcal{O}}}=\sum_i (\bar{\mathcal{O}}_i^2+ \sigma_i^2) p_i - \bar{\mathcal{O}}^2,
\end{equation}
where $\sigma_i$ is the statistical error of $\bar{O}_i$.  For our
final values, we separately quote our statistical and systematic
error, namely,
\begin{gather}
	(\sigma^{\rm stat})^2 = \sum_i \sigma_i^2 p_i
	\quad \text{and} \quad
	(\sigma^{\rm sys})^2 = \sum_i \bar{\mathcal{O}}_i^2 p_i - \bar{\mathcal{O}}^2,
\end{gather}
respectively.

\subsection{$Q^2$-dependence of the strange form factors}
In what follows, we  discuss  the fits to the $Q^2$-dependence of
the strange electric and magnetic form factors,
$G^s_E(Q^2)$ and $G^s_M(Q^2)$, starting with the dipole and
Galster-like forms, followed by the $z$-expansion.

\subsubsection{Galster-like and dipole analysis}

\begin{figure}
	\centering
	\includegraphics[width=1\linewidth]{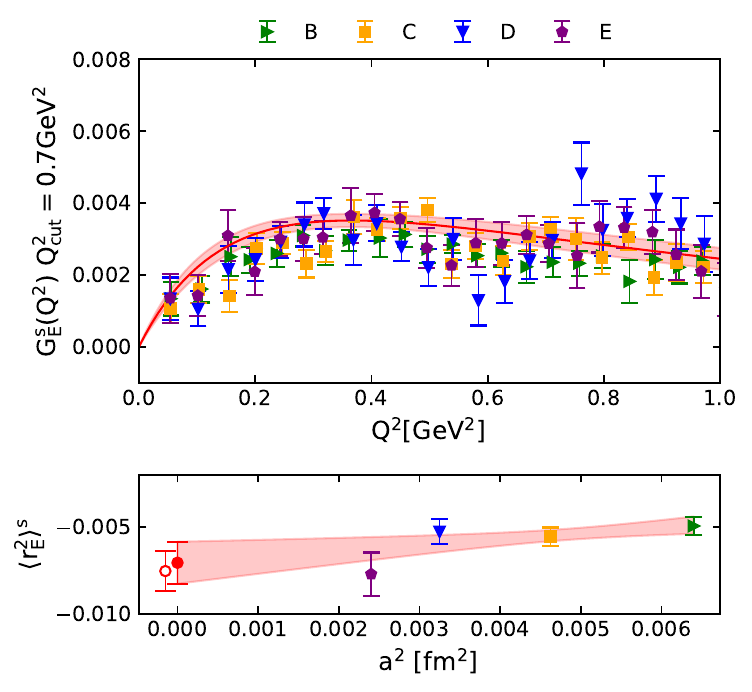}
	\caption{Results for the strange electric $G^s_E(Q^2)$ form
	factor from the \texttt{cB211.072.64} (right-pointing green
	triangles), \texttt{cC211.060.80} (orange squares),
	\texttt{cD211.054.96} (downward-pointing blue triangles), and
	\texttt{cE211.044.112} (purple pentagons) ensembles. The red
	band in the top panel corresponds to the continuum limit
	form factor using a one-step Galster-like fit performed
	simultaneously to all ensembles with momentum cut, $Q^2_{\rm
		cut} = 0.7\;\rm GeV^2$. The bottom panel shows the
	continuum limit result for the strange electric radius
	($\langle r^2_E\rangle^s$), with the four filled points at
	four values of the lattice spacing corresponding to
	Galster-like fits to each ensemble, the filled red point and
	the red band to the linear extrapolation in $a^2$, and the
	open red symbol denoting the one-step continuum limit
	result.}
	\label{fig:1step_2step_GE_comparison}
\end{figure}

As discussed, we use the Galster-like parameterization for the electric form factor, constraining it  to be zero at $Q^2=0$. In
Fig.~\ref{fig:1step_2step_GE_comparison}, we show our data for
$G_E^s(Q^2)$ for $Q^2\in[0,1] \; \rm GeV^2$. The different symbols
denote the results for the four different ensembles analyzed in this
work. As mentioned in
Sec.~\ref{sec:extraction}, including non-zero momenta in the sink
increases the available $Q^2$ values from $\mathcal{O}(20)$ to
$\mathcal{O}(300)$ for each ensemble for $Q^2\in[0,1.3)$. Thus, for a
clearer visualization of our data, we bin the form factor values
over $Q^2$ intervals. The fits are
performed on error weighted averages in $250$ bins of width $0.004\;\rm GeV^2$, averaging over very closely spaced $Q^2$ points, thereby reducing statistical noise.  We perform both one- and
two-step Galster-like fits to the data of the four ensembles. The
filled red point denotes the continuum limit result obtained using
linear extrapolation in $a^2$  from the
two-step fitting procedure. The slightly shifted open red point
corresponds to the result from the one-step approach. As can be
seen, the two approaches yield compatible results. In
Table~\ref{tab:ges_galster}, we tabulate the results  when varying the maximum value of the momentum transfer $Q^2$ used in the fit, $Q^2_{\rm cut}$. We also include the reduced
$\tilde{\chi}^2\equiv\chi^2/N_\mathrm{dof}$ value for each fit, where $N_\mathrm{dof}$
are the number of degrees of freedom.

\begin{table}
	\caption{Continuum limit results on the strange electric charge radius $\langle r^2_{\rm
		E} \rangle^s$ and reduced $\tilde{\chi}^2\def\chi^2/{N_{dof}}$ obtained when using one-step Galster-like fits.}
	\label{tab:ges_galster}
	\begin{tabular}{ccc}
		\hline
		\hline
		$Q^2_{\rm cut} [\rm GeV^2]$ & $\langle r_E^2 \rangle^{s}$ [fm$^2$] & $	\tilde{\chi}^2$ \\
		\hline
		0.40                        & -0.0060(17)                          & 0.9310           \\
		0.50                        & -0.0056(12)                          & 0.9870           \\
		0.70                        & -0.0075(12)                          & 1.1310           \\
		0.85                        & -0.00709(95)                         & 1.2340           \\
		1.00                        & -0.00665(82)                         & 1.4640           \\
		\hline
	\end{tabular}
\end{table}

\begin{figure}
	\centering
	\includegraphics[width=1\linewidth]{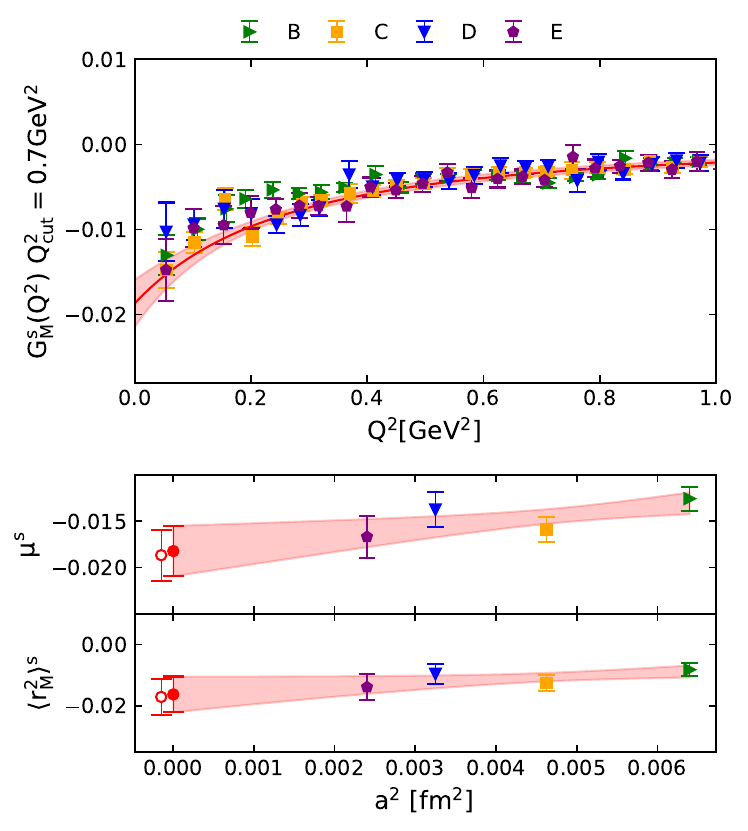}
	\caption{Results for the strange magnetic $G^s_M(Q^2)$ form
	factor, with both one- and two-step dipole fits with
	momentum cut $Q^2_{\rm cut} = 0.7\;\rm GeV^2$. The notation
	is the same as in Fig.~\ref{fig:1step_2step_GE_comparison}.
	}
	\label{fig:1step_2step_GM_comparison}
\end{figure}

A similar analysis is performed for the magnetic form factor
$G_M^s(Q^2)$ using the dipole Ansatz. The results are  shown in Fig.~\ref{fig:1step_2step_GM_comparison}.

The results are tabulated in Table~\ref{tab:gms_dipole}, where we see
that the one- and two-step approaches yield consistent results at the
continuum limit. We proceed with the one-step approach in what
follows, since it is directly accessible for the model averaging
procedure and the fit takes into account the complete covariance of
the parameters.

\begin{table}
	\caption{The strange magnetic moment ($\mu^s$), magnetic radius
	($\langle r^2_{\rm M} \rangle^s$), and  reduced $\tilde{\chi}^2$
	extracted from one-step dipole fits to the strange magnetic form factor
	for each ensemble as we increase $Q^2_{\rm cut}$.}
	\label{tab:gms_dipole}
	\begin{tabular}{cccc}
		\hline
		\hline
		$Q^2_{\rm cut} [\rm GeV^2]$ & $\mu^s$     & $\langle r_M^2 \rangle^{s}$ [fm$^2$] & $\tilde{\chi}^2$ \\
		\hline
		0.40                        & -0.0145(37) & -0.0068(65)                          & 1.0950           \\
		0.50                        & -0.0169(31) & -0.0132(65)                          & 1.1620           \\
		0.70                        & -0.0187(28) & -0.0170(60)                          & 1.3040           \\
		0.85                        & -0.0191(26) & -0.0181(55)                          & 1.4230           \\
		1.00                        & -0.0200(23) & -0.0201(52)                          & 1.5020           \\
		\hline
	\end{tabular}
\end{table}

\subsubsection{$z$-expansion analysis}
\begin{figure}[h]
	\centering
	\includegraphics[width=\linewidth]{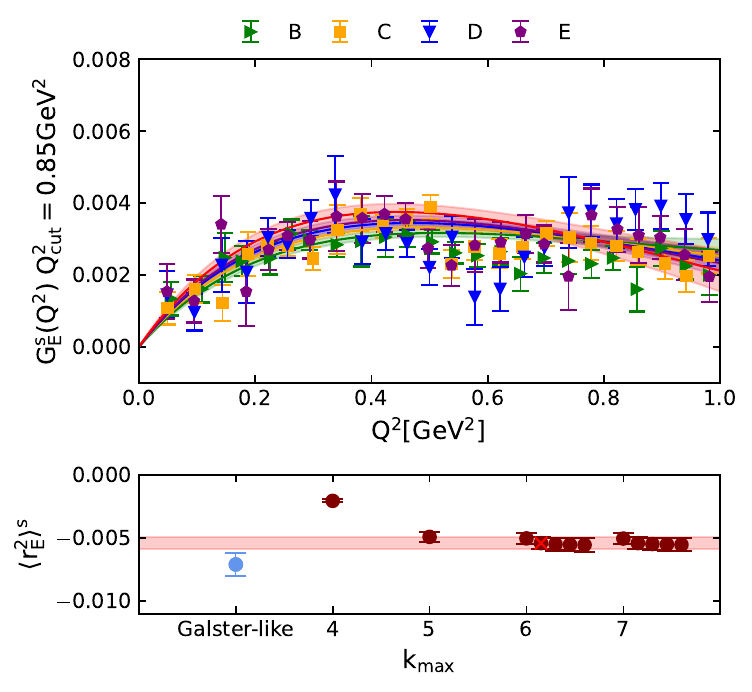}
	\caption{The $z$-expansion fit to $G^s_E(Q^2)$ with $Q_{\rm
		cut}^2=0.85$~GeV$^2$ (top panel) and the dependence of the
		extracted radius on the $z$-expansion order ($k_{\rm max}$)
		and prior widths ($w$) (bottom panel). In the top panel, the
		green, orange, blue, and purple bands show the result of the
		one-step $z$-expansion fit for $a=0.080$, 0.068, 0.057, and
		0.049~fm, respectively and the red band for $a=0$. In the
		bottom panel, for each set of points at $k_{\rm max}=6$ and
		7, the leftmost point corresponds to the result when using
		$w$=1 in Eq.~(\ref{eq:zexp_priors}) and with $w$ increasing
		by 2 up to $w=9$ for the rightmost point. For $k_{\rm
		max}=4$ and~5 no priors are employed, as explained in the
		text. For the bands in the top panel, we use $k_{max}=6$ and
		$w=3$, indicated by the red cross and red band in the bottom
		panel. The blue point in the bottom panel plot is the result
		from the one-step Galster-like fit, shown for comparison.}
	\label{fig:zexp_ges}
\end{figure}

For $z$-expansion fits we opt to use only the one-step approach, since
a two-step approach in which we fit individually each ensemble becomes
unstable for relatively small values of $k_{\rm max}$ and thus does
not allow for a robust analysis of the stability of the $z$-expansion
as $k_{\rm max}$ increases.
For the electric form factor, we set the zeroth order parameters of
the $z$-expansion, namely $c_{0,0}$ and $c_{0,2}$, to zero in order to
impose $G_E^s(0)=0$ (see Eq.~(\ref{eq:zexp_a2})). For the first order
parameters, that directly relate to the radius, and those of the next
order, we do not use any priors, whereas for all higher order
parameters we allow for Gaussian priors with widths as in
Eq.~\ref{eq:zexp_priors}.  In the top panel of
Fig.~\ref{fig:zexp_ges}, we show results for the four ensembles.  With
the bands, we show the resulting $Q^2$-fits for each ensemble,
obtained by using the value of the lattice spacings of the respective
ensemble in Eq.~\ref{eq:zexp_a2}. The red band denotes the continuum
limit when using $Q^2_{\rm cut} = 0.7\;\rm GeV^2$.

The robustness of the $z$-expansion fit, and in particular of the
extracted radius, as the order ($k_{\rm max}$) and prior widths ($w$
in Eq.~\ref{eq:zexp_priors}) in the $z$-expansion are increased is
investigated and the results shown in Fig.~\ref{fig:zexp_ges}. Namely,
we investigate how the continuum limit result for the strange electric
radius ($\langle r^2_E \rangle^s$) varies as a function of $k_{\rm
max}$ for the width $w$ of the prior in the range $1\leq w \leq 9$.
We note that since for fits with $k_{\rm max}=4,\;5$ we do not use any
priors, only one point is presented. We observe convergence at $k_{\rm
max}=6,\; w=3$, as can be seen in Fig.~\ref{fig:zexp_ges} and we
choose this value for our final result. This choice of order and width
is adopted for all momenta cuts and the results are summarized in
Table~\ref{tab:ges_z-exp} together with the corresponding reduced
$\tilde{\chi}^2$ of each fit.

\begin{table}
	\caption{The strange electric charge radius $\langle r^2_{\rm E}
		\rangle^s$ obtained using the one-step approach and the
	$z$-expansion with $k_{\rm max}=6$ and $w=3$ as we  increase
	$Q^2_{\rm cut}$. The values of  the reduced $\tilde{\chi}^2$ are also provided.}
	\label{tab:ges_z-exp}
	\begin{tabular}{ccc}
		\hline
		\hline
		$Q^2_{\rm cut} [\rm GeV^2]$ & $\langle r_E^2 \rangle^{s}$ [fm$^2$] & $	\tilde{\chi}^2$ \\
		\hline
		0.40                        & -0.0052(10)                          & 0.9050           \\
		0.50                        & -0.00483(80)                         & 0.9730           \\
		0.70                        & -0.00564(58)                         & 1.1150           \\
		0.85                        & -0.00540(48)                         & 1.2250           \\
		1.00                        & -0.00540(41)                         & 1.4550           \\
		\hline
	\end{tabular}
\end{table}

\begin{figure}
	\centering
	\includegraphics[width=\linewidth]{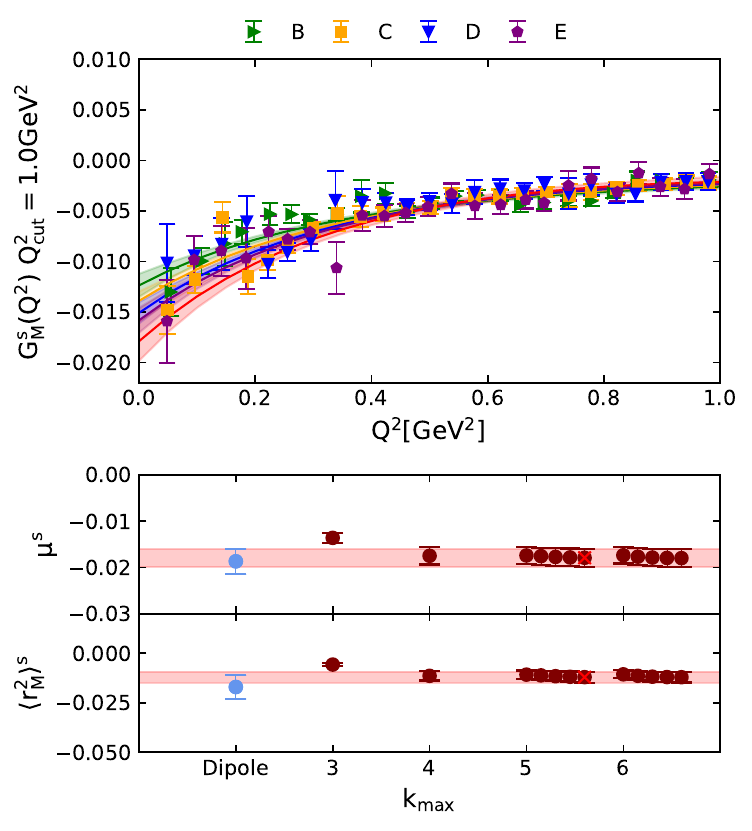}
	\caption{Results on $G^s_M(Q^2)$ using the $z$-expansion with
		$Q_\mathrm{cut}^2=1.0\;\rm GeV^2$ and $k_{max}=5$ and $w=9$, (top
		panel) and the dependence of the extracted magnetic moment and
		radius on the $z$-expansion order ($k_{\rm max}$) and prior widths
		($w$) (bottom panel). In the top panel, the green, orange, blue, and
		purple bands show the result of the one-step $z$-expansion fit for
		the B, C, D and E ensembles, respectively and the red band
		corresponds to the continuum limit using the red crossed point in bottom panels. In the bottom panel, we show
		results on the magnetic moment $\mu^s$ and square radius when
		using the $z$-expansion with $k_{\rm max}=3, 4, 5,$ and 4, and with
		$w$=1 for the leftmost points and increasing by 2 up to $w=9$. The red cross and the red band running in the bottom panel corresponds to the final result chosen.
		The blue point is the result from the
		one-step dipole fit, shown for comparison.}
	\label{fig:zexp_gms}
\end{figure}

For the strange magnetic form factors, when using the $z$-expansion,
we use non-zero coefficients for $k=0$. We use no priors for the
parameters of the first two orders of the $z$-expansion, which yield
the magnetic moment and the magnetic radius, as well as the next
order. Thus, priors are only employed for coefficients corresponding
to $k\geq3$. In the top panel of Fig.~\ref{fig:zexp_gms}, results for
the strange magnetic form factors for all ensembles are plotted along
with a continuum limit band resulting from the one-step
$z$-expansion. We show the results of the investigation of the
dependence on $k_{\rm max}$ and prior widths $w$ of the magnetic
moment ($\mu^s$) and the magnetic radius ($\langle r^2_M\rangle^s$) in
the bottom panel of Fig.~\ref{fig:zexp_gms}.

While a stronger dependence on $w$ is observed compared to the
electric case, convergence is achieved for $w=9$, with consistent
agreement among the results for $k_{\rm max}=$4, 5, and 6. Henceforth,
we take $k_{\rm max}=5$ and $w=9$, as denoted by the red cross in the
bottom panel, thereby allowing the same number of fit parameters as
the electric case for $k_{\rm max}=6$. We perform fits for various
momenta cuts and summarize the obtained radius and moment in
Table~\ref{tab:gms_z-exp}.

\begin{table}
	\caption{The strange magnetic moment ($\mu^s$), magnetic radius
	($\langle r^2_{\rm M} \rangle^s$), and  $\tilde{\chi}^2$
	obtained using the one-step $z$-expansion with
	$k_{\rm max}=5$ and $w=9$ as  we increase $Q^2_{\rm cut}$.}
	\label{tab:gms_z-exp}
	\begin{tabular}{cccc}
		\hline
		\hline
		$Q^2_{\rm cut} [\rm GeV^2]$ & $\mu^s$     & $\langle r_M^2 \rangle^{s}$ [fm$^2$] & $ \tilde{\chi}^2$ \\
		\hline
		0.40                        & -0.0135(29) & -0.0033(49)                          & 1.0540            \\
		0.50                        & -0.0153(26) & -0.0073(44)                          & 1.1250            \\
		0.70                        & -0.0151(24) & -0.0073(39)                          & 1.2810            \\
		0.85                        & -0.0169(22) & -0.0103(33)                          & 1.4060            \\
		1.00                        & -0.0179(20) & -0.0121(28)                          & 1.4890            \\
		\hline
	\end{tabular}
\end{table}

\subsection{Charm electromagnetic form factor}
Similarly to the strange quark contribution, the charm electromagnetic
form factors can provide a better understanding of the sea quark
dynamics in the nucleon. A non-zero value of the charm electromagnetic
form factors can provide insights into the asymmetry of the charm
anti-charm distributions inside the nucleon. However, the heavy mass of
the charm quark means that this would require energy exceeding the
hadronic energy scale, potentially leading to a suppressed or consistent with zero
contribution within error bars.
\begin{figure}[h!]
	\centering
	\includegraphics[width=\linewidth]{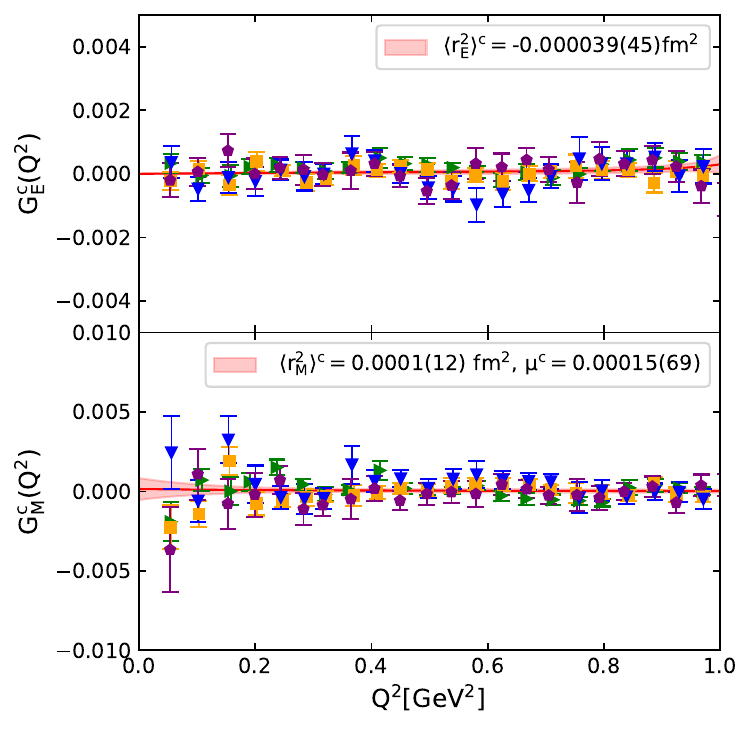}
	\caption{Results for $G^c_E(Q^2)$ and $G^c_M(Q^2)$ using the
		Galster-like and dipole Ans\"atze  without $a^2$ dependence
		and with $Q_\mathrm{cut}^2=0.7$~GeV$^2$. The green, orange,
		blue and purple points show the results for the four
		ensembles used. The legend shows results for the magnetic
		moment ($\mu^c$) and square electric ($\langle
			r^2_E\rangle^c$) and magnetic radii ($\langle
			r^2_M\rangle^c$) with fits as explained in the text.}
	\label{fig:gem_c}
\end{figure}
Indeed, using the same procedure as done for the strange
contribution, we obtain results for the charm electric ($G_E^c(Q^2)$)
and magnetic form factors ($G_M^c(Q^2)$) that are
compatible with zero. A Galster-like (dipole form) fit without $a^2$
dependence to the electric (magnetic) form factors also results in
values of the fit parameters consistent with zero. We show our data
and fits in Fig.~\ref{fig:gem_c}.
It should be noted that only one other lattice QCD study by $\chi$QCD Collaboration obtains a non-zero electromagnetic charm form factor for the nucleon~\cite{Sufian:2020coz}.
They use 17 different quark masses with overlap valence fermions computed on three ensembles of \Nf{2}{1} RBC/UKQCD domain wall fermion configurations with $m_{\pi}\in\{139, 300, 330\}$. Using two-state fit to the summed ratio of three- to two-point functions to extract the form factors at each $Q^2$, their final results are obtained by employing a simultaneous fit using z-expansion parameterization which incorporates chiral, continuum and infinite volume extrapolations.
However, we would like to stress that their results are compatible with ours within the statistical errors.

\section{Results and comparison to other studies}
\label{sec:final_results}

\subsection{Final results}

To obtain our final values for the strange electric and magnetic radii and
magnetic moment, we quantify the systematic error from the fit Ans\"atze
and momentum transfer cuts using the model averaging procedure
explained in Sec.~\ref{sec:MA}.  The models used are summarized in the
first column of Table~\ref{tab:final_results}, and include one-step
dipole for the magnetic case, one-step Galster-like for the electric
case, and one-step $z$-expansion fits for both cases, with varying
values of $Q^2_{\rm cut}$. We provide the strange electric radius,
strange magnetic radius, and the strange magnetic moment of the
nucleon along with the probability associated with each fit using the
AIC.

\begin{table*}
	\centering
	\caption{Results for the nucleon strange electric and magnetic
		radii squared and for the magnetic moment with the
		corresponding probabilities for each fit as obtained in the
		model-average. Here, Ansatz II corresponds to the dipole
		form for the magnetic form factor and Galster-like for the
		electric form factor.}
	\label{tab:final_results}
	\begin{tabular}{c|cc|ccc}
		\hline
		\hline  $Q^2_{\rm cut} [\rm GeV^2]$ & $\langle r^2_{\rm E} \rangle^s$ [fm$^2$] & prob(\%) & $\mu^s$           & $\langle r^2_{\rm M} \rangle^s$ [fm$^2$] & prob(\%) \\
		\hline
		z-exp 0.40                          & -0.0052(10)                              & 0.0000   & -0.0135(29)       & -0.0033(49)                              & 0.0000   \\
		z-exp 0.50                          & -0.00483(80)                             & 0.0000   & -0.0153(26)       & -0.0073(44)                              & 0.0000   \\
		z-exp 0.70                          & -0.00564(58)                             & 0.0000   & -0.0151(24)       & -0.0073(39)                              & 0.0000   \\
		z-exp 0.85                          & -0.00540(48)                             & 0.9744   & -0.0169(22)       & -0.0103(33)                              & 0.0000   \\
		z-exp 1.00                          & -0.00540(41)                             & 0.0000   & -0.0179(20)       & -0.0121(28)                              & 0.9918   \\
		\hline
		Ansatz II 0.40                      & -0.0060(17)                              & 0.0000   & -0.0145(37)       & -0.0068(65)                              & 0.0000   \\
		Ansatz II 0.50                      & -0.0056(12)                              & 0.0000   & -0.0169(31)       & -0.0132(65)                              & 0.0000   \\
		Ansatz II 0.70                      & -0.0075(12)                              & 0.0000   & -0.0187(28)       & -0.0170(60)                              & 0.0000   \\
		Ansatz II 0.85                      & -0.00709(95)                             & 0.0256   & -0.0191(26)       & -0.0181(55)                              & 0.0000   \\
		Ansatz II 1.00                      & -0.00665(82)                             & 0.0000   & -0.0200(23)       & -0.0201(52)                              & 0.0082   \\
		\hline Final results                & -0.00545(49)(26)                         &          & -0.01792(195)(18) & -0.01212(280)(72)                        &          \\
		\hline
	\end{tabular}
\end{table*}

\begin{figure}[h!]
	\centering
	\includegraphics[width=1\linewidth]{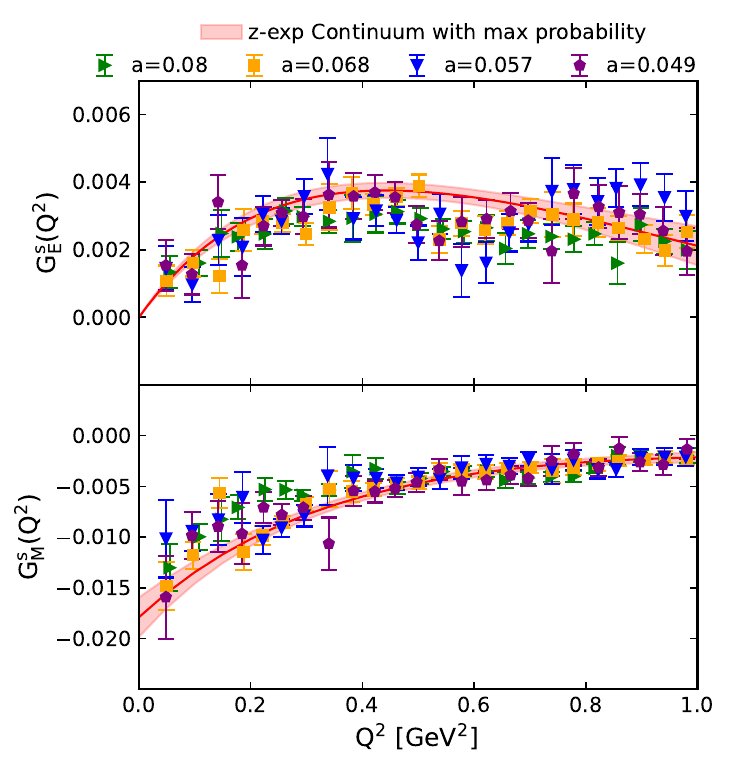}
	\caption{Results for the strange electric form factor,
		$G^s_E(Q^2)$ (top) and strange magnetic form factor,
		$G^s_M(Q^2)$ (bottom) as a function of $Q^2$ for the four
		ensembles analyzed with the corresponding lattice spacings
		indicated in the header of the figure. The red band
		corresponds to the most probable continuum limit fit results
		obtained using the one-step $z$-expansion (see
		Table~\ref{tab:final_results}).}
	\label{fig:GEMs}
\end{figure}

\begin{figure}[h!]
	\centering
	\includegraphics[width=0.475\textwidth]{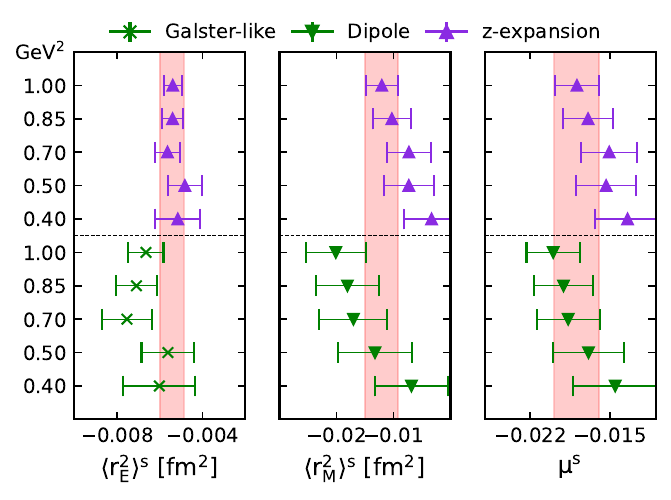}
	\caption{Values of $\langle r^2_E\rangle^s$, $\langle
		r^2_M\rangle^s$ and $\mu^s$ obtained as a result of fitting
	to $G_E^s$ and $G_M^s$ with different $Q^2_{\rm cut}$. The
	upward-pointing violet triangles denote results from the
	$z$-expansion fits, green crosses denote the Galster-like
	fit, and downward-pointing green triangles denote the
	results from dipole fits. The vertical red band
	corresponds to the model-averaged value with combined
	statistical and systematic errors using AIC. }
	\label{fig:Results_Q2}
\end{figure}

As can be seen from the values of the  probabilities,
the one-step $z$-expansion parameterization using $Q^2_{\rm cut} =
	0.7\;\rm GeV^2$ dominates for both electric and magnetic form factors. In
Fig.~\ref{fig:GEMs}, we plot the continuum limit of the electric
and magnetic form factors as well as the fit results for the most probable model. Most other models are compatible
within $1\sigma$ of the most probable model, which can be visualized in
Fig~\ref{fig:Results_Q2}, where we plot our results for the radii and
magnetic moments for varying $Q^2_{\rm cut}$ for the one-step approach
for all parameterizations.

\subsection{Comparison with other studies}

\begin{figure}
	\centering
	\includegraphics[width=0.5\textwidth]{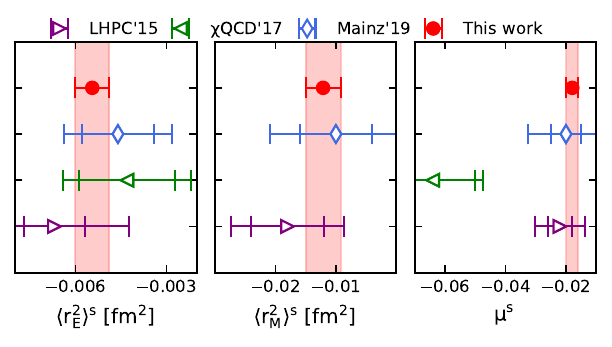}
	\caption{Results for the strange electric and magnetic radii and the
		magnetic moments of the nucleon obtained within this
		work (red circles) where statistical and systematical errors are combined using the AIC procedure. The red vertical bands correspond to our value with combined
		statistical and systematic error. We compare to previous results by LHPC (right-pointing open triangle)~\cite{Green:2015wqa}, $\chi$QCD (left-pointing open triangle)~\cite{Sufian:2016pex} and the Mainz group (blue diamond)~\cite{Djukanovic:2019jtp}. In all these case the smaller
		error corresponds to their statistical error and the larger
		error is the statistical and systematic error added in quadrature. }
	\label{fig:Compare_lattice_results}
\end{figure}

In Fig.~\ref{fig:Compare_lattice_results}, we show our results for the
strange electric and magnetic square radii and strange magnetic moment
to other lattice QCD studies. Because of the small magnitude of the
strange contribution and the stochastic error related to the
estimation of the strange quark loop, the number of lattice results
for these quantities are limited. We also restrict our comparison to
the most recent results by the respective collaborations. We summarize
the works to which we compare to below.

\begin{itemize}

	\item Results from our previous
	      analysis~\cite{Alexandrou:2019olr} use the coarsest
	      physical point ensemble of the current analysis,
	      \texttt{cB211.072.64}. We observe very good agreement for
	      the strange electric and magnetic radii, $\langle r^2_E
		      \rangle^s = -0.0048(6)\;\rm fm^2$, $\langle r^2_M \rangle^s
		      = -0.015(9)\;\rm fm^2$ and the strange magnetic moment,
	      $\mu^s=-0.017(4)$, indicating cut-off effects smaller than
	      our statistical errors for these quantities.

	\item The Mainz group~\cite{Djukanovic:2019jtp} obtains
	      results in the continuum limit using six \Nf{2}{1} CLS
	      ensembles with all heavier-than-physical $m_{\pi}\in(200,
		      360)\;\rm MeV$.  They primarily employ the summation method and
	      compare with plateau fits in order to estimate  excited
	      state systematics. They use the $z$-expansion to parameterize their $Q^2$
	      dependence and extract   the radii and moments for each
	      ensemble. This is followed by a
	      continuum and chiral extrapolation using heavy baryon chiral
	      perturbation theory. We observe good
	      agreement within their quoted statistical uncertainties for
	      the strange magnetic radius and moment and agreement within
	      their combined uncertainties for the strange electric
	      radius.

	\item The $\chi$QCD Collaboration presents continuum limit
	      results using a mixed setup with overlap valence fermions
	      computed on \Nf{2}{1} RBC/UKQCD domain wall fermion
	      configurations. The continuum limit has been obtained using
	      four ensembles with $m_{\pi}\in(135, 350)\;\rm MeV$, which
	      includes one physical point ensemble, and a total of 24 valence quark
	      masses on the four ensembles. In order to remove
	      excited-state contamination, they performed two-state fits
	      in combination with both the ratio method and the
	      summed-ratio method.  Additionally, the $z$-expansion is used
	      for their $Q^2$-dependence parameterization and their final
	      results are extracted using a chiral extrapolation to the
	      physical point.  We compare with their final results for the
	      strange electric radius and the strange magnetic moment of
	      the nucleon in
	      Fig.~\ref{fig:Compare_lattice_results}.
	      We observe very good agreement
	      for the electric radius, whereas their magnetic moment
	      is smaller as compared to all lattice QCD results.

	\item In a publication demonstrating the improvement obtained
	      by using hierarchical probing, the results by the LHPC
	      collaboration~\cite{Green:2015wqa}, used \Nf{2}{1}
	      clover-improved Wilson action with $m_{\pi} = 317\; \rm MeV$
	      ensemble at $a\sim0.114 \;\rm fm$ and volume
	      $32^3\times96$. They use a source-sink time separation of $1.14
			\;\rm fm$ in order to obtain results with sufficiently
	      suppressed excited state contamination. Additionally, they
	      use the $z$-expansion with  $t_{\rm cut}=(2m_{\pi})^2$ to obtain final results. We find that
	      our results are in very good agreement for both electric and
	      magnetic radii and the magnetic moment.

\end{itemize}

\begin{figure}
	\centering
	\includegraphics[width=\linewidth]{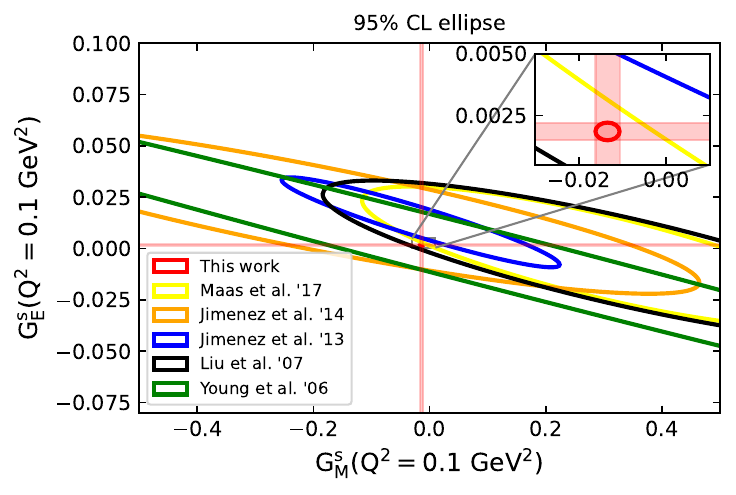}
	\caption{Comparison of the $95\%$ confidence level curves for
		the electric and magnetic strange form factors at
		$Q^2=0.1\;\rm GeV^2$. The red vertical and horizontal bands
		corresponds to our results for $G_E^s(Q^2=0.1\;\rm GeV^2)$
		and $G_M^s(Q^2=0.1\;\rm GeV^2)$ with the red ellipse in the
		inset corresponding to the confidence level ellipse. The
		remaining ellipses correspond to the reanalyses of the
		experimental data with yellow taken
		from~\cite{Maas:2017snj}, orange
		from~\cite{Gonzalez-Jimenez:2014bia}, blue
		from~\cite{Gonzalez-Jimenez:2011qkf}, black
		from~\cite{Liu:2007yi} and green from~\cite{Young:2006jc}.}
	\label{fig:95_cl}
\end{figure}

In addition to the comparison with the lattice QCD results, we also
present a comparison with various results obtained using the
reanalyses of the experimental data of parity violating asymmetry
measurements. In Fig.~\ref{fig:95_cl}, we show the $95\%$ confidence
ellipses for the strange electric and magnetic form factors at
$Q^2=0.1\;\rm GeV^2$ provided by various
studies~\cite{Maas:2017snj,Gonzalez-Jimenez:2014bia,Gonzalez-Jimenez:2011qkf,Liu:2007yi,Young:2006jc}. As
can be observed, the uncertainty associated with these measurements is
indeed very large owing to the indirect measurement through the
asymmetry cross section, as opposed to the very precise value we
obtain, denoted by the red ellipse in the inset figure. The tilt in
the ellipse for the experimental result can be attributed to the
correlation between the measurements of  $G_E(Q^2)$ and $G_M(Q^2)$
which is not observed in the lattice correlation functions used in our
results.

\section{Conclusions}
\label{sec:conclusions}

We present the  analysis of the nucleon strange electromagnetic
form factors using four ensembles of \Nf{2}{1}{1} twisted mass
fermions at four lattice spacings spanning $a\in[0.08, 0.049]$
simulated at approximately physical value of the pion mass. This allows, for the first time, to  take the continuum limit without any chiral extrapolation.
The statistics of the three-point functions are increased, by using two boosted frames in addition to the lab frame, giving access $\mathcal{O}(300)$ values of the momentum
transfer squared in $Q^2\in[0,1.3)\;\rm GeV^2$, thereby allowing for a very
precise $Q^2$-dependence analysis.

We combine the $Q^2$-dependence and continuum limit in $a^2$ fits to obtain results at the continuum
limit (one-step procedure).   For the strange electric form factors we employ both
one-step Galster-like and $z$-expansion to extract the electric radii,
whereas for the magnetic form factor we use one-step dipole and
$z$-expansion.  We include systematics arising from the Ansatz
dependence and the momentum transfer cuts used in our fitting
procedure. Our final results, thus, include   statistical and
systematic errors, obtained from a
model-averaging procedure.

Our values on the form factors are significantly more precise
when compared with the experimental results, providing meaningful constraints on  the strange electric and magnetic
radii and the strange magnetic moment of the nucleon. In summary, our
final values for these quantities are,
\begin{gather}
	\langle r^2_E\rangle^s = -0.00545(49)(26) \;\;\rm fm^2, \\
	\langle r^2_M\rangle^s = -0.01212(280)(72) \;\;\rm fm^2,\\
	\mu^s = -0.01792(195)(18).
\end{gather}

A similar  analysis of  the charm
electromagnetic form factors yields results that are
consistent with zero at the statistical precision of our data.

\section*{Acknowledgments}

We thank Matteo Di Carlo, Antonio Evangelista, Roberto Frezzotti,
Giuseppe Gagliardi and Vittorio Lubicz, for useful discussions and
crosschecks on the analysis of the renormalization factors.  C.A.,
S.B., G.K., and G.S.  acknowledge partial support by the projects
Baryon8 (POSTDOC/0524/0001), MuonHVP (EXCELLENCE/0524/0017), PulseQCD
(EXCELLENCE/0524/0269), DeNuTra (EXCELLENCE/0524/0455), IMAGE-N
(EXCELLENCE/0524/0459), HyperON (VISION ERC-PATH 2/0524/0001),
StrongILA (EXCELLENCE/0524/0001) and partonWF (VISION ERC/0525/0010)
co-financed by the European Regional Development Fund and the Republic
of Cyprus through the Research and Innovation Foundation, as well as
the European Joint Doctorate AQTIVATE that received funding from the
European Union’s research and innovation program under the Marie
Sklodowska-Curie Doctoral Networks action, Grant Agreement No
101072344. C.A acknowledges support by the University of Cyprus
projects ``Nucleon-GPDs'' and ``PDFs-LQCD''. This project received
funding from the European Research Council (ERC) via the project
"LEEX" grant agreement 101170304. Funded by the European Union. Views
and opinions expressed are however those of the author(s) only and do
not necessarily reflect those of the European Union or the European
Research Council Executive Agency (ERCEA). Neither the European Union
nor the ERCEA can be held responsible for them.  B.P. is supported by
ENGAGE which received funding from the EU's Horizon 2020 Research and
Innovation Programme under the Marie Skłodowska-Curie GA
No. 101034267. This work was supported by grants from the Swiss
National Supercomputing Centre (CSCS) under project with s1174. The
authors gratefully acknowledge the Gauss Centre for Supercomputing
e.V. (www.gauss-centre.eu) for funding this project by providing
computing time through the John von Neumann Institute for Computing
(NIC) on the GCS Supercomputers JUWELS~\cite{JUWELS} and JUWELS
Booster~\cite{JUWELS-BOOSTER} at J\"ulich Supercomputing Centre
(JSC). The authors acknowledge access to the JUPITER supercomputer,
which is funded by the EuroHPC Joint Undertaking, the German Federal
Ministry of Research, Technology and Space, and the Ministry of
Culture and Science of the German state of North Rhine-Westphalia,
through the JUPITER Research and Early Access Program (JUREAP) as part
of the EuroHPC project application EHPC-EXT-2023E02-052. The authors
also acknowledge the Texas Advanced Computing Center (TACC) at
University of Texas at Austin for providing HPC resources.

\bibliography{refs}

\appendix

\section{Decomposition of lattice matrix elements}
\label{sec:appendix_equations}

In this appendix, we summarize the Euclidean space equations for
the ratio in Eq.~\ref{eq:full_ratio} when considered for the case of
$p'\neq 0$. For compact representation, we take $G_E \equiv G_E(Q^2)$,
$G_M \equiv G_M(Q^2)$, $\Pi_{\mu,0} \equiv
	\Pi^\mu(\Gamma_0;\vec{p}\,', \vec{p}) $, $\Pi_{\mu,k} \equiv
	\Pi^\mu(\Gamma_k;\vec{p}\,', \vec{p})$ where ${\vec p}$ (${\vec p}'$)
corresponds to the initial (final) momentum. The four momenta are then
given by $\{E,\vec{p}\}$ ($\{E',\vec{p'}\}$), where $E$ ($E'$) is the
initial (final) energy.

\begin{align}
	\Pi_{\mu,0} & =
	-iCG_E\Big[ \left(p'_\mu + p_\mu\right) \left(m\left( E'+ E + m \right) - p'_\rho p_\rho \right) \Big]
	\nonumber       \\&\quad
	+ \frac{CG_M}{2m} \Big[\delta_{\mu 0}( 4m^2 + Q^2 )(m^2 + p'_\rho p_\rho) - i E Q^2 p'_\mu
	\nonumber       \\&\quad
		+ 2 i m^2 (E' - E)(p'_\mu - p_\mu) - i E' Q^2p_\mu
	\nonumber       \\&\quad
		- i m Q^2 (p'_\mu + p_\mu) (2m^2 + Q^2 + 2p'_\rho p_\rho) \Big]
	\label{eq:pi_full_0}
\end{align}

\begin{align}
	\Pi_{\mu,k} & = CG_E\Big[ \epsilon_{\mu k 0 \rho} (p_\rho' - p_\rho) (m^2 - p_\sigma' p_\sigma)
	\nonumber                                                                                       \\&\quad
		- i\epsilon_{\mu k \rho \sigma} p_\rho' p_\sigma (E'  + E)+\epsilon_{\mu 0 \rho \sigma}p_\rho' p_\sigma (p_k' + p_k) \Big]
	\nonumber                                                                                       \\&\quad
	+ \frac{ CG_M}{2m} \Big[ m\epsilon_{\mu k 0 \rho} (p_\rho' - p_\rho) (2m^2 + Q^2 + 2p'_\sigma p_\sigma)
	\nonumber                                                                                       \\&\quad
		+ 2im\epsilon_{\mu k \rho \sigma} p'_\rho p_\sigma(2m+E' + E +\frac{Q^2}{2m})
	\nonumber                                                                                       \\&\quad
		-2m\epsilon_{\mu 0 \rho \sigma}p'_\rho p_\sigma (p_k' + p_k) \Big],
	\label{eq:eq:pi_full_k}
\end{align}
where $C$ is a kinematic factor given by
\begin{equation}
	C = \frac{m(4m^2+Q^2)^{-1}}{E (E'+m)} \sqrt{\frac{E (E'+m)}{E'(E+m)}}
\end{equation}

\section{Tabulated results}
\label{sec:appendix_results}
The results of the calculated form factors are presented for each ensemble in this section. In Table~\ref{tab:results_B}, Table~\ref{tab:results_C}, Table~\ref{tab:results_D}, and Table~\ref{tab:results_E}, we provide our values for the strange electric and magnetic form factors for \texttt{cB211.072.64}, \texttt{cC211.060.80}, \texttt{cD211.054.96}, and \texttt{cE211.044.112} ensembles respectively. Since these are  disconnected contributions with additional sink momenta leading to $O(300)$ $Q^2$ values for the
all ensembles, we show results averaged into 25 equal bins within the interval $Q^2 = [0,1.0)
	\;\rm GeV^2$. More precisely, for bin $i$ starting from $Q^2_i$ and
ending at $Q^2_{i+1}=Q^2_{i}+\delta Q^2$ we average all form factor
values with $Q^2_i\le Q^2 < Q^2_{i+1}$ weighted by their errors.

We additionally provide the fit parameters and
their associated covariance matrix for the most probable fit to the
$Q^2$-dependence of the form factors, i.e., those used to plot the
bands in Fig.~\ref{fig:GEMs} in Table~\ref{tab:fitp}.

\begin{table}[H]
	\centering
	\caption{Results for the strange electric and magnetic form factors for the \texttt{cB211.072.64} ensemble.}
	\label{tab:results_B}
	\begin{tabular}{|c|c|c|}
		\hline
		\hline
		$Q^2$  & $G_E^s$     & $G_M^s$      \\
		\hline
		0.0550 & 0.00134(47) & -0.0130(23)  \\
		0.1080 & 0.00161(37) & -0.0100(13)  \\
		0.1479 & 0.00248(69) & -0.0083(22)  \\
		0.1771 & 0.00238(41) & -0.0071(12)  \\
		0.2256 & 0.00246(45) & -0.0053(11)  \\
		0.2628 & 0.00315(39) & -0.00535(89) \\
		0.2933 & 0.00298(27) & -0.00587(54) \\
		0.3403 & 0.00284(33) & -0.00536(63) \\
		0.3826 & 0.00293(70) & -0.0036(16)  \\
		0.4239 & 0.00305(54) & -0.0033(10)  \\
		0.4626 & 0.00313(35) & -0.00513(61) \\
		0.5042 & 0.00292(32) & -0.00423(55) \\
		0.5442 & 0.00261(30) & -0.00354(47) \\
		0.5822 & 0.00253(32) & -0.00345(49) \\
		0.6221 & 0.00265(39) & -0.00349(58) \\
		0.6562 & 0.00203(47) & -0.00439(77) \\
		0.6968 & 0.00247(41) & -0.00418(61) \\
		0.7389 & 0.00239(36) & -0.00406(49) \\
		0.7785 & 0.00231(38) & -0.00402(53) \\
		0.8158 & 0.00247(34) & -0.00324(46) \\
		0.8567 & 0.00160(62) & -0.00199(88) \\
		0.8986 & 0.00274(49) & -0.00198(61) \\
		0.9410 & 0.00225(38) & -0.00225(47) \\
		0.9815 & 0.00200(55) & -0.00194(74) \\
		\hline
	\end{tabular}
\end{table}

\begin{table}[H]
	\centering
	\caption{Results for the strange electric and magnetic form factors for the \texttt{cC211.060.80} ensemble.}
	\label{tab:results_C}
	\begin{tabular}{|c|c|c|}
		\hline
		\hline
		$Q^2$  & $G_E^s$     & $G_M^s$      \\
		\hline
		0.0490 & 0.00108(46) & -0.0148(23)  \\
		0.0965 & 0.00160(39) & -0.0118(13)  \\
		0.1440 & 0.00122(50) & -0.0056(15)  \\
		0.1875 & 0.00258(62) & -0.0115(17)  \\
		0.2215 & 0.00283(37) & -0.00984(97) \\
		0.2566 & 0.00280(27) & -0.00851(62) \\
		0.2992 & 0.00246(33) & -0.00675(67) \\
		0.3412 & 0.00325(68) & -0.0052(17)  \\
		0.3811 & 0.00367(46) & -0.0055(10)  \\
		0.4199 & 0.00334(31) & -0.00531(65) \\
		0.4610 & 0.00355(27) & -0.00474(54) \\
		0.5016 & 0.00389(33) & -0.00465(62) \\
		0.5381 & 0.00230(40) & -0.00347(74) \\
		0.5771 & 0.00279(37) & -0.00346(61) \\
		0.6199 & 0.00259(41) & -0.00358(69) \\
		0.6595 & 0.00278(35) & -0.00327(55) \\
		0.7012 & 0.00315(35) & -0.00313(52) \\
		0.7405 & 0.00304(68) & -0.0035(12)  \\
		0.7776 & 0.00289(48) & -0.00313(78) \\
		0.8218 & 0.00280(42) & -0.00281(64) \\
		0.8589 & 0.00264(35) & -0.00250(54) \\
		0.9053 & 0.00232(43) & -0.00236(57) \\
		0.9417 & 0.00197(45) & -0.00209(59) \\
		0.9810 & 0.00253(49) & -0.00225(64) \\
		\hline
	\end{tabular}
\end{table}

\begin{table}[H]
	\centering
	\caption{Results for the strange electric and magnetic form factors for the \texttt{cD211.054.96} ensemble.}
	\label{tab:results_D}
	\begin{tabular}{|c|c|c|}
		\hline
		\hline
		$Q^2$  & $G_E^s$     & $G_M^s$      \\
		\hline
		0.0485 & 0.00145(65) & -0.0101(38)  \\
		0.0954 & 0.00097(51) & -0.0094(19)  \\
		0.1424 & 0.00228(75) & -0.0083(25)  \\
		0.1855 & 0.00208(86) & -0.0061(25)  \\
		0.2220 & 0.00305(53) & -0.0103(14)  \\
		0.2553 & 0.00285(39) & -0.00908(89) \\
		0.2959 & 0.00358(51) & -0.0079(11)  \\
		0.3378 & 0.0042(11)  & -0.0039(29)  \\
		0.3843 & 0.00293(63) & -0.0042(13)  \\
		0.4247 & 0.00315(46) & -0.00419(93) \\
		0.4613 & 0.00288(39) & -0.00463(75) \\
		0.4999 & 0.00221(50) & -0.00406(86) \\
		0.5387 & 0.00306(58) & -0.00428(97) \\
		0.5777 & 0.00139(78) & -0.0031(12)  \\
		0.6215 & 0.00161(61) & -0.00280(91) \\
		0.6631 & 0.00250(56) & -0.00300(79) \\
		0.6977 & 0.00277(52) & -0.00231(71) \\
		0.7394 & 0.00373(99) & -0.0033(15)  \\
		0.7780 & 0.00378(73) & -0.0024(10)  \\
		0.8215 & 0.00343(69) & -0.00332(90) \\
		0.8543 & 0.00382(57) & -0.00332(75) \\
		0.8980 & 0.00392(63) & -0.00213(79) \\
		0.9410 & 0.00354(70) & -0.00209(85) \\
		0.9790 & 0.00299(75) & -0.00209(87) \\
		\hline
	\end{tabular}
\end{table}

\begin{table}[H]
	\centering
	\caption{Results for the strange electric and magnetic form factors for the \texttt{cE211.044.112} ensemble.}
	\label{tab:results_E}
	\begin{tabular}{|c|c|c|}
		\hline
		\hline
		$Q^2$  & $G_E^s$     & $G_M^s$      \\
		\hline
		0.0482 & 0.00154(76) & -0.0159(41)  \\
		0.0949 & 0.00127(60) & -0.0099(23)  \\
		0.1416 & 0.00341(79) & -0.0090(25)  \\
		0.1845 & 0.00154(95) & -0.0097(30)  \\
		0.2231 & 0.00270(58) & -0.0071(15)  \\
		0.2557 & 0.00309(39) & -0.0078(10)  \\
		0.2942 & 0.00298(52) & -0.0071(12)  \\
		0.3400 & 0.00363(97) & -0.0107(26)  \\
		0.3843 & 0.00358(68) & -0.0055(15)  \\
		0.4224 & 0.00369(51) & -0.0055(11)  \\
		0.4587 & 0.00355(46) & -0.00521(85) \\
		0.4972 & 0.00274(54) & -0.00461(94) \\
		0.5377 & 0.00226(58) & -0.0033(10)  \\
		0.5779 & 0.00282(73) & -0.0045(13)  \\
		0.6225 & 0.00291(56) & -0.00439(92) \\
		0.6655 & 0.00314(53) & -0.00390(87) \\
		0.6970 & 0.00286(52) & -0.00421(79) \\
		0.7389 & 0.00196(94) & -0.0025(15)  \\
		0.7784 & 0.00366(74) & -0.0018(11)  \\
		0.8231 & 0.00327(63) & -0.00315(86) \\
		0.8599 & 0.00310(78) & -0.0013(11)  \\
		0.8969 & 0.00304(59) & -0.00261(89) \\
		0.9386 & 0.00256(71) & -0.00286(98) \\
		0.9814 & 0.00195(70) & -0.00136(98) \\
		\hline
	\end{tabular}
\end{table}

\begin{table*}
	\centering
	\caption{Fit parameters for the most probable fits shown in
		Fig.~\ref{fig:GEMs} at $a=0$. For each form factor indicated in the
		first column, the $Q^2_\mathrm{cut}$ and fit form of the most
		probable fit is provided in the second column and the symbols used
		in the text for the fit parameters are provided in the third
		column. The values and errors of the fit parameters are given in the
		fourth column and their corresponding covariance matrix in the
		fifth. }
	\label{tab:fitp}
	\begin{tabular}{|c|c|c|c|c|}
		\hline
		\hline
		                                                  & Fit & Fit parameters, $\vec{p}$ & Results for $\vec{p}$ & Covariance of fit parameters, cov$(\vec{p})$ \\
		\hline
		$G_E^s$                                           &

		z-exp, $Q^2_{\rm cut} = 0.85\;\rm GeV^2$          &

		$\begin{pmatrix}
				 c_{1,0} \\ c_{2,0}\\ c_{3,0} \\ c_{4,0}
			 \end{pmatrix}$           &

		$\begin{pmatrix}
				 0.0897(86) \\ -0.64(13) \\ 1.15(79) \\ 0.1(2.2)
			 \end{pmatrix}$   &

		$\begin{pmatrix}
				 0.0001  & -0.0010 & 0.0035  & -0.0033 \\
				 -0.0010 & 0.0171  & -0.0866 & 0.1240  \\
				 0.0035  & -0.0866 & 0.6192  & -1.3564 \\
				 -0.0033 & 0.1240  & -1.3564 & 4.6568  \\
			 \end{pmatrix}$                                                                                                                     \\

		\hline

		$G_M^s$                                           &

		z-exp, $Q^2_{\rm cut} = 1.0\;\rm GeV^2$           &

		$\begin{pmatrix}
				 c_{0,0} \\ c_{1,0}\\ c_{2,0} \\ c_{3,0}
			 \end{pmatrix}$           &

		$\begin{pmatrix}
				 -0.0179(20) \\ 0.196(45) \\ -0.78(39) \\ 1.1(1.5)
			 \end{pmatrix}$ &

		$\begin{pmatrix}
				 0.000004  & -0.000080 & 0.000504  & -0.001142 \\
				 -0.000080 & 0.002028  & -0.015768 & 0.046742  \\
				 0.000504  & -0.015768 & 0.151048  & -0.553670 \\
				 -0.001142 & 0.046742  & -0.553670 & 2.391708  \\
			 \end{pmatrix}$                                                                                                             \\

		\hline
	\end{tabular}
\end{table*}

\end{document}